\documentclass[format=acmsmall, screen,nonacm]{acmart}

\acmSubmissionID{Insert-paper-id-here (ACM ICFP 2020)}
\setcopyright{none}

\usepackage{pgfplotstable,multicol,caption}
\usepackage{amsmath,amssymb,amsfonts}
\usepackage{algorithmic}
\usepackage{graphicx}
\usepackage{textcomp}
\usepackage{xcolor}
\usepackage{comment}
\usepackage{wrapfig}
\usepackage[utf8]{inputenc}
\usepackage[english]{babel}
\usepackage{subcaption}
\graphicspath{ {images/} }
\usepackage{hyperref}
\usepackage{glossaries}
\usepackage{xspace}
\usepackage[inline]{enumitem}
\usepackage{tikz}
\usepackage{import}

\glsdisablehyper
\def\BibTeX{{\rm B\kern-.05em{\sc i\kern-.025em b}\kern-.08em
		T\kern-.1667em\lower.7ex\hbox{E}\kern-.125emX}}

\usepackage{listings}
\usepackage{amssymb}

\newcommand{\canFlowTo}{\ensuremath{\sqsubseteq}}
\newcommand{\liojoin}{\ensuremath{\sqcup}}
\newcommand{\liomeet}{\ensuremath{\sqcap}}

\def\withcolor{}

\ifdefined\withcolor
  \definecolor{fstarblue}{rgb}{0.0, 0.0, 1.0}
  \definecolor{haskellstr}{rgb}{0.2, 0.2, 0.6}
  \definecolor{haskellred}{rgb}{1.0, 0.0, 0.0}
  \definecolor{gray_ulisses}{gray}{0.55}
  \definecolor{castanho_ulisses}{rgb}{0.59,0.42,0.15}
  \definecolor{preto_ulisses}{rgb}{0.55,0.28,0.59}
  \definecolor{green_ulises}{rgb}{0.59,0.42,0.15}
\else
	\definecolor{fstarblue}{gray}{0.1}
	\definecolor{haskellstr}{gray}{0.1}
	\definecolor{haskellred}{gray}{0.1}
	\definecolor{gray_ulisses}{gray}{0.1}
	\definecolor{castanho_ulisses}{gray}{0.1}
	\definecolor{preto_ulisses}{gray}{0.1}
	\definecolor{green_ulisses}{gray}{0.1}
\fi

\def\codesize{\small}

\lstdefinelanguage{HaskellUlisses} {
	basicstyle=\ttfamily\codesize,
	sensitive=true,
	morecomment=[s][\color{gray_ulisses}\ttfamily\itshape\codesize]{(*}{*)},
	morecomment=[l][\color{gray_ulisses}\ttfamily\itshape\codesize]{//},
	morestring=[b]",
	stringstyle=\color{haskellstr},
	basewidth={0.53em},
	showstringspaces=false,
	numberstyle=\codesize,
	numberblanklines=true,
	showspaces=false,
	breaklines=true,
	showtabs=false,
	tabsize=4,
    literate={ {/\\}{{$\land$}}2
             {->}{{$\rightarrow$}}2
             {<=}{{$\leq$}}2
             {>=}{{$\geq$}}2
             {forall}{{$\forall$}}1
			 {'a}{{$\alpha$}}1
			 {'b}{{$\beta$}}1
			 {'l}{{$l$}}1
             {fun}{{$\lambda$}}1
             {True}{{$\top$}}1
             {False}{{$\bot$}}1
             {~int}{{$\mathbb{Z}$}}1
             {~nat}{{$\mathbb{N}$}}1
             {==>}{{$\Rightarrow$}}1
             {=>}{{$\Rightarrow$}}1
             {`eq`}{{=}}2
             {`equals`}{{=}}2
             {`bottom`}{{$\bot$}}2
             {`meet`}{{$\liomeet$}}2
             {`join`}{{$\liojoin$}}2
             {`canFlow`}{{$\canFlowTo$}}2
			 {`notCanFlow`}{{$\not \canFlowTo$}}2
			 {`hole`}{{$\bullet$}}1
           },
	emph=
	{[1] TERM, Tot, Type, bool, Lemma, ensures, requires, Ifc, IFC, IfcClearance, GlobalInt, GTot
	},
	emphstyle={[1]\color{fstarblue}},
	emph=
	{[2] class, module, instance, match, with, if, then, else, let, rec, type, val, in, effect,noeq, private
	},
	emphstyle={[2]\color{castanho_ulisses}},
	emph=
	{[3]
		lattice, value, equals, canFlow, meet, join, bottom, top,
		hasEraser, 
		simple_lattice,
        lawBot, lawFlowReflexivity, lawFlowAntisymetry, lawFlowTransitivity, 
        lawReflexivity, lawAntisymetry, lawTransitivity, 
        lawMeet, lawJoin, labels, 
        lt, lmeet, ljoin, lcanFlow, eq,
        labeled, labeledTCB
	},
	emphstyle={[3]\color{preto_ulisses}\textbf},
	emph=
	{[4]
        Low, Medium, High
	},
	emphstyle={[4]\color{green_ulises}\textbf},
	emph=
	{[5] assume, admit, admitP
	},
	emphstyle=[5]\color{red}\textbf,
	emph={[6] leq, equals, join', c_0, c_1
	},
	emphstyle=[6]\color{green}\textbf,
}

\lstnewenvironment{code}
{\lstset{language=HaskellUlisses}}
{}

\lstnewenvironment{scode}
{\lstset{language=HaskellUlisses,basicstyle=\ttfamily\footnotesize,keepspaces,mathescape}}
{}

\lstnewenvironment{mcode}
{\lstset{language=HaskellUlisses,columns=fullflexible,keepspaces,mathescape}}
{}

\lstnewenvironment{ccode}
{\lstset{language=C,columns=fullflexible,keepspaces,mathescape}}
{}

\lstMakeShortInline[language=HaskellUlisses,mathescape,keepspaces,mathescape,basicstyle=\ttfamily\codesize,breakatwhitespace]|

\newcommand{\tstar}{$^{\star}$}
\newcommand{\liostar}{\texorpdfstring{LIO\tstar}{LioStar}}
\newcommand{\fstar}{\texorpdfstring{F\tstar}{FStar}}
\newcommand{\metastar}{\texorpdfstring{Meta\tstar}{MetaStar}}
\newcommand{\datastar}{\texorpdfstring{DB\tstar}{DBStar}}
\newcommand{\busstar}{\texorpdfstring{BUS\tstar}{BUSStar}}\newcommand{\mmustar}{\texorpdfstring{MMU\tstar}{MMUStar}}
\newcommand{\lowstar}{\texorpdfstring{Low\tstar}{LowStar}}

\newcommand{\ie}{\textit{i.e.,}\xspace}
\newcommand{\eg}{\textit{e.g.,}\xspace}

\definecolor{mGreen}{rgb}{0,0.6,0}
\definecolor{mGray}{rgb}{0.5,0.5,0.5}
\definecolor{mPurple}{rgb}{0.58,0,0.82}
\definecolor{backgroundColour}{rgb}{0.95,0.95,0.92}
\lstdefinestyle{CStyle}{
	commentstyle=\color{mGreen},
	keywordstyle=\color{magenta},
	numberstyle=\tiny\color{mGray},
	stringstyle=\color{mPurple},
	basicstyle=\footnotesize,
	breakatwhitespace=false,         
	breaklines=true,                 
	captionpos=b,                    
	keepspaces=true,                 
	numbers=left,                    
	numbersep=5pt,                  
	showspaces=false,                
	showstringspaces=false,
	showtabs=false,                  
	tabsize=2,
	language=C
}

\newcommand\fNI{\ensuremath{\text{fNI}}\xspace}
\newcommand\bind[2]{\ensuremath{(#1\text{:}#2)}}
\newcommand\erase[1]{\ensuremath{\epsilon_{\texttt{#1}}}\xspace}
\newcommand\eraseapp[2]{\ensuremath{\erase{#1}(#2)}}
\newcommand\eval[3]{\ensuremath{\downarrow\!\!^{\texttt{#3}}{\texttt{#1}(\texttt{#2})}}\xspace}

\setacronymstyle{long-short}

\newacronym{lio}{LIO}{Labeled Input Output}

\newacronym{krml}{KreMLin}{KreMLin}
\newacronym{fstar}{\fstar}{FStar}

\newacronym{liostar}{\liostar}{Labeled Input Output in \fstar}
\newacronym{ifc}{IFC}{Information Flow Control}
\newacronym{iot}{IoT}{Internet of Things}
\newacronym{lcp}{LCP}{Label Creep Problem}
\newacronym{lowstar}{\lowstar}{LowStar}
\newacronym{metastar}{\metastar}{MetaStar}
\newacronym{c}{C}{C programming language}
\newacronym{gcc}{GCC}{GNU C Compiler}
\newacronym{os}{OS}{Operating System}
\newacronym{libos}{libOS}{Library Operating System}
\newacronym{ast}{AST}{Abstract Syntax Tree}

\newacronym{api}{API}{Application Programming Interface}
\newacronym{tcb}{TCB}{Trusted Computing Base}
\newacronym{linux}{linux}{linux}
\newacronym{wasm}{WASM}{Web Assembly}
\newacronym{asm}{ASM}{Assembly}

\newacronym{mmu}{MMU}{Memory Management Unit}

\newacronym{mmustar}{MMU\tstar}{Memory Management Unit in \fstar}
\newacronym{datastar}{\datastar}{database in \fstar}
\newacronym{wp}{Weakest Precondition}{WP}
\newacronym{busstar}{\busstar}{BUSStar}
\newacronym{ipc}{IPC}{Inter-Process Communication}
\newacronym{edr}{EDR}{Event Data Recorder}
\newacronym{haskell}{Haskell}{}
\newacronym{lop}{LoF}{Lines of \fstar}
\newacronym{loc}{LoC}{Lines of Code}
\newacronym{noc}{PPT}{Percentage of Primitives' Time}

\newcommand{\lioDeian}{D\liostar\xspace}
\newcommand{\lioStatic}{G\liostar\xspace}
\newcommand{\lioHybrid}{S\liostar\xspace}

\pgfplotsset{compat=1.8}
\pgfplotstableread{
	paradigm        lio     lioS    lioC
	{\datastar} 10.571      9.229     9.602
	{\busstar}   7.144      3.238     5.610
	{\mmustar}  16.179      8.450     12.449
}\mytable

\pgfplotstableread{
	paradigm        lio     lioS    lioC
	{\datastar}    456      414     426
	{\busstar}     100      68      92
	{\mmustar}     192      155     176
}\locdata

\begin{document}
	
	\title{\liostar: Low Level Information Flow Control with \fstar}

	\author{Jean-Joseph Marty}
		\affiliation{
		\institution{INRIA, IRISA}
		\city{Rennes}
		\country{France}}
	\email{jean-joseph.marty@inria.fr}
	\author{Lucas Franceschino}
		\affiliation{
		\institution{INRIA, IRISA}
		\city{Rennes}
		\country{France}}
	\email{lucas.franceschino@inria.fr}
	\author{Jean-Pierre Talpin}
		\affiliation{
		\institution{INRIA, IRISA}
		\city{Rennes}
		\country{France}}
	\email{jean-pierre.talpin@inria.fr}
	\author{Niki Vazou}
	\affiliation{
		\institution{IMDEA Software Institute}
		\city{Madrid}
		\country{Spain}}
	\email{niki.vazou@imdea.org}

	
	\begin{abstract}
We present \gls{liostar}, a verified framework that enforces 
information flow control (IFC) policies developed in \acrshort{fstar}
and automatically extracted to C.
We encapsulated IFC policies into effects, 
but using \acrshort{fstar} we derived efficient, low-level, 
and provably correct code. 
Concretely, runtime checks are lifted to static proof obligations, 
the developed code is automatically extracted to C and 
proved non-interferent using metaprogramming.  
We benchmarked our framework on three clients and observed up to 54\%
speedup when IFC runtime checks are proved statically. 
Our framework is designed to aid development of embedded devices 
where both enforcement of security policies and low-level efficient code is 
critical. 
\end{abstract}

	\maketitle
	\glsresetall

\section{Introduction}
\label{sec:intro}

Low-level embedded devices are part of connected environments 
that bring wisdom into systems
(\eg smart cars, appliances, and houses)
or combine various sources of information, as in connected health. 
Such devices are programmed in low-level languages, like C,
to form components, \eg bus or GPS, that need   
to be efficient and resource-constrained. 
At the same time, being part of interconnected systems, 
it is critical that they enforce security policies for
component separation. 

\gls{ifc}~\citep{Sabelfeld:2006} policies can be used to ensure component separation, 
but the current techniques that enforce such policies 
either use heavy runtime checks~\citep{Austin12,Austin17,Tromer16,Yang16}
or rely on advanced type systems of high level programming languages~\citep{Schoepe14,Myers19,buiras2015hlio}.
For example, \gls{lio} relies on Haskell's monads to soundly and effectively
enforce \gls{ifc} policies when reading/writing to databases or to the 
web~\citep{stefan:2017:flexible,Parker:2019:LIF:3302515.3290388}. 

Direct application of \gls{lio} to embedded devices faces two major obstacles.  
First, \gls{lio}'s policy enforcement relies on automatically generated 
runtime checks that would unpredictably crash the device. 
Second, Haskell's garbage collector and lazy evaluation might 
lead to memory leaks rendering automatic code extraction for 
devices with limited memory resources nearly impossible. 

In this work we develop \gls{liostar}, a variant of \gls{lio}
that is implemented and verified in \gls{fstar}~\citep{mumon},
and automatically extracted to efficient C via \gls{krml}~\citep{lowstar}.
Concretely, we used \gls{fstar}'s effects to encode \gls{ifc} encapsulation 
in the spirit of \gls{lio}'s monadic programming. 
This lead to three advantages: 
First, we lifted policy enforcement from runtime checks 
to static proof obligations, using \gls{fstar}'s Dijkstra Monads~\citep{Maillard19} 
leading, when possible (\S~\ref{discussion:crashfreedom}), to provably crash-free code.
Second, the low-level code generated from \gls{krml} does not require a 
runtime library nor a garbage collector, 
so it is suitable to execute on embedded devices.
Third, we used \gls{fstar}'s meta-programming support \gls{metastar}~\citep{metafstar} 
to prove that the actual \gls{liostar}'s clients enjoy non-interference. 
In short, we propose a methodology to generate both verified
and runtime optimized \gls{ifc} applications. 

Our contributions are the following: 
\begin{itemize}
	\item We developed \gls{liostar}, an \gls{fstar} library that enforces  
	\gls{ifc} policies (\S~\ref{sec:implementation}). 
	\gls{liostar} is statically verified by \gls{fstar} which allows for both
	1) more efficient implementations, since dynamic \gls{ifc} checks are 
	statically proved and are thus redundant (\S~\ref{subsec:implementation:static}) and 
	2) low-level generated C code that is automatically extracted using \gls{krml}. 
	\item We designed a mechanized meta-programming procedure that proves that clients of \gls{liostar} 
	are non-interferent and applied it to two benchmarks and various toy clients
	(\S~\ref{section:metathery}).
	Concretely, we used \gls{fstar}'s meta-programming facilities (\gls{metastar})
	to prove, for the first time, non-interference of the actual \gls{ifc} clients, 
	instead of idealized models.
	We report conclusions and some limitations of this endeavor.
	\item We benchmarked \gls{liostar} on three client applications (\S~\ref{sec:benchmarks}):
	1) \acrshort{busstar}, that implements a bus system with \gls{ifc} policies between communicating system components, 
	2) \acrshort{mmustar}, that implements a software-defined memory management unit with policies on concurrent resources, and 
	3) \acrshort{datastar}, that implements a database with explicit \gls{ifc} policies. 	
	From these benchmarks, we conclude that 
	extraction of C code under the \gls{ifc} effect is possible 
	and that the replacement of runtime checks with static proofs leads to 
	cleaner code and up to 54\% speedup (Table \ref{table:benchmark_speed}).
\end{itemize}


	\section{Overview}
\label{sec:overview}
\label{discussion:crashfreedom}

\begin{wrapfigure}{r}{0.4\textwidth}
	\begin{center}
	  \includegraphics[width=0.38\textwidth]{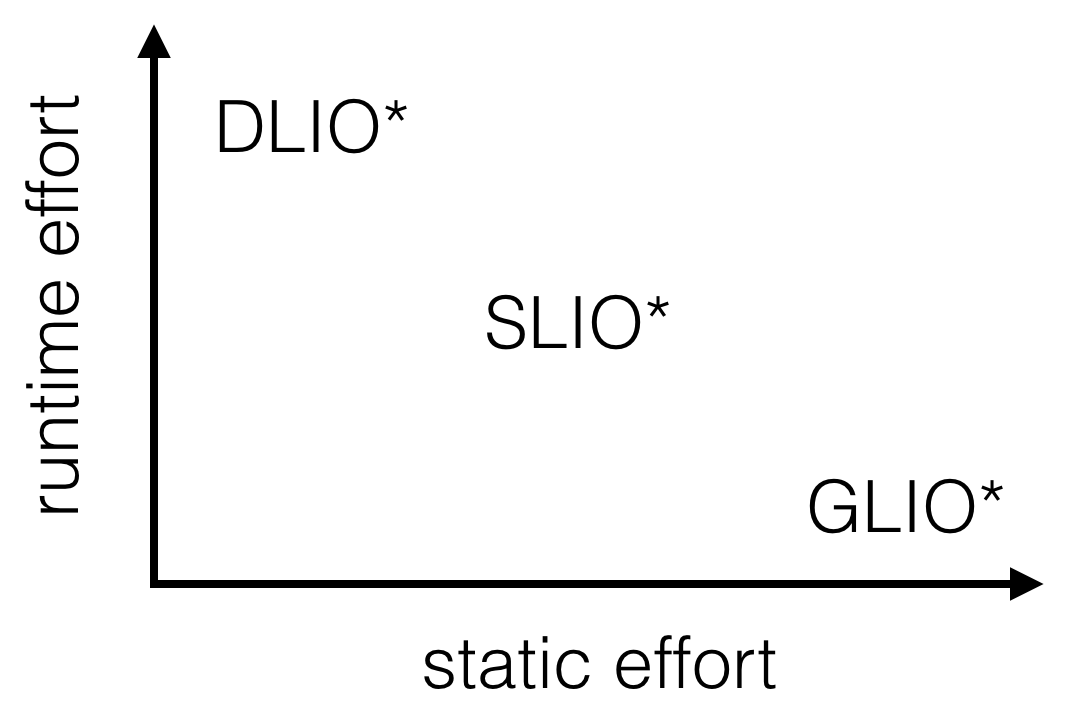}
	\end{center}
	\caption{The three versions of LIO*}
	\label{fig:diagram}
\end{wrapfigure}

This section provides an overview of the interface of three 
\gls{ifc} libraries we implemented, that are summarized in 
Figure~\ref{fig:diagram}.
\lioDeian, presented in~\S~\ref{subsec:overview:dynamic}, 
provides \gls{ifc} using runtime checks, which may unpredictably 
raise exceptions during execution, but requires zero 
static verification effort from the developers. 
\lioHybrid, presented in~\S~\ref{subsec:overview:static}, 
lifts \gls{ifc} to the verification of static proof obligations, 
thus minimizing runtime overhead and failures. 
However, the developers are required to statically discharge 
all proof obligations. 
\lioStatic, presented in~\S~\ref{subsec:overview:ghost},  
both lifts all \gls{ifc} checks as static proof obligations 
and marks all \gls{ifc} privilege tracking information as 
computationally irrelevant to remove all \gls{ifc}-related information 
from the runtime code, making it ideal for closed-loop execution 
but inconvenient when dynamic \gls{ifc} information is necessary.
We conclude that in applications where performance is critical, 
\eg embedded systems, the static effort required by \lioStatic
might be worth the effort, but for mainstream software development 
\lioHybrid provides an ideal trade-off among runtime checks and 
verification effort. 

\subsection{\lioDeian: naive translation of \gls{lio} into \gls{fstar}}
\label{subsec:overview:dynamic}

We start with an overview of the interface of \lioDeian, an \gls{ifc} library 
that, like \gls{lio}, we uses a lattice of 
security labels to protect sensitive data. 
 
\paragraph{Labels to protect values}
We use security labels to express information flow control policies. 
As a trivial security label, consider an enumeration of three values:
|type label = $\mid$ Low $\mid$ Medium $\mid$ High|.
To protect data, we assign them labels and ensure that data labeled as,
\eg |Medium| can only be accessed by users with |Medium| or higher privileges.

The label system can be generalized to any lattice~\citep{Denning:1976}
where 
the privileges hierarchy is defined by a partial order relation \canFlowTo.
The label interface is generalized as an \gls{fstar} type class 
defined in~\S~\ref{subsec:implementation:labels}. 
Importantly, 
the |lattice| type class contains the least upper bound (|join|, \liojoin) 
of two labels, the label ordering (|canFlowTo|, \canFlowTo), 
as well as proofs that the above methods form a partial order. 

\paragraph{The IFC Effect}

\lioDeian enforces \gls{ifc} using runtime checks. 
Concretely, it provides two methods to label and unlabel data, 
that are defined in~\S~\ref{subsec:implementation:dynamic} 
and have the interface below. 
\begin{mcode}
  val label	  : 'a -> 'l -> Ifc (labeled 'a)
  val unlabel : labeled 'a -> Ifc 'a
\end{mcode}
The method |label| takes a value |v| 
and a label |l| as inputs and returns a |labeled| value, \ie 
it wraps the value |v| with the label |l|. 
Dually, |unlabel| returns the value of its |labeled|, protected input. 
Both operations have an |Ifc| effect, which tracks the 
\textit{current label} |cur|, that is, the accumulated privileges 
required to perform the current computation. 
Concretely, |unlabel lv| updates the current label by joining it 
with the label of |lv|.
|label v l| only returns a labeled value when the current label 
can flow to |l| (|cur| \canFlowTo |l|).
%
When it cannot, we are at risk of transmitting a value |v| of 
high privilege (\ie |cur|) to a level of lower security (\ie |l|). 
To prevent this violation, |label| fails with an exception.

\paragraph{Notations}
To simplify the type signatures, we use |'a|
for types that support decidable equality |'a:Type{hasEq 'a}|
and |'l| for instances of the |lattice| type class, 
a simplification permitted by using \gls{fstar}'s functors 
(as explained in~\S~\ref{subsec:implementation:labels}). 
Additionally, most trivial |ensures| and |requires| clauses, 
\eg |requires fun _ -> True|, are omitted from function signatures 
to improve readability.

\paragraph{Example} \label{section:overview:eqlabeled}
Consider below two functions that label and unlabel data below.
On the left, |eqLabeled| takes two labeled data 
as input and unlabels them to compare their values. 
On the right, |checkLabeled| labels its input value
and uses it to call |eqLabeled|.

\noindent
\begin{minipage}{0.4\textwidth}
\begin{mcode}
  let eqLabeled (v1 v2 :labeled 'a) 
                : Ifc bool
    let i1 = unlabel v1 in
    // cur := cur `join` labelOf v1 
    let i2 = unlabel v2 in i1 = i2
    // cur := cur `join` labelOf v2
\end{mcode}
\end{minipage} %
\begin{minipage}{0.6\textwidth}
\begin{mcode}
  let checkLabeled (l:'l) (i:'a) (lv:labeled 'a)
				   : Ifc bool 
    let lv' = label i l in
    // exception if cur `notCanFlow` l 
    eqLabeled lv lv'
    // cur := cur `join` labelOf lv 
\end{mcode}
\end{minipage} %
Each time |eqLabeled| unlabels a value,
the current label is joined with the label of 
the value, while when |checkLabeled| labels a value
a runtime check is performed. 


\paragraph{Extraction to C}
\lioDeian, as described so far, is equivalent to \gls{lio}
where Haskell's |LIO| monad is replaced by \gls{fstar}'s 
|Ifc| effect. 
\citet{Wadler2003} established that, in theory, the transition from 
monads to effects is always possible. In practice, 
this transition from Haskell to \gls{fstar} gives us 
the ability to extract low-level C code, using  
\gls{krml}~\citep{lowstar} that automatically translates 
an \gls{fstar} program to readable C code.
%
%
For example, if our main program calls the function 
|checkLabeled(l,i)| with 
|label l| and |int32 i|, \gls{krml} will generate 
the following C code. 
\begin{mcode}
  bool checkLabeled__int32_t(label l, int32_t i, labeled___int32_t lv) {
	label cur = getCurrent();
	if (!canFlow(cur, l)) fail("invalid labelling");
	labeled___int32_t lv_ = { .data = i, .tag = l };
	return eqLabeled__int32_t(lv, lv_);
  }  
\end{mcode}
That is, the C code directly accesses the current label |cur|
and checks if it can flow to the argument label |l|. 
If so, it packs the input value |i| with the label |l|
and uses it to call the C translation of the |eqLabeled| function. 
Otherwise, it fails with a runtime check. 
Below is the extraction of the |eqLabeled| function
instantiated on integer labeled values. 
\begin{mcode}
  bool eqLabeled(labeled___int32_t v1, labeled___int32_t v2) {
	labels c0 = getCurrent();
	labels c1 = join(c0, labelOf__int32_t(v1));
	setCurrent(c1);
	int32_t i1 = v1.data;
	labels c2 = getCurrent();
	labels c3 = join(c2, labelOf__int32_t(v2));
	setCurrent(c3);
	int32_t i2 = v2.data;
	return i1 == i2;
  }
\end{mcode}
That is, for each call to |unlabel v|, the C translation 
1/ gets the current label, 2/ joins it with the label of |v|, 
3/ sets the current label to the joint label, and 4/ gets the value of |v|. 
Finally, the comparison of the two values is returned. 

This translation from \lioDeian to low-level C
brings us one step closer to our goal of generating 
an \gls{ifc} library suitable to program embedded devices. 
But, we still face a big challenge: 
the generated code contains runtime checks (as seen in the above example)
that are generated by calls to |label| and are 
%
not documented 
and thus unpredictable by \lioDeian's clients.
These potential runtime failures make \lioDeian unsuitable for our goal, 
as 
hardware devices have critical sections that must not fail.


\subsection{\lioHybrid: turning \gls{ifc} runtime checks into static proof obligations}
\label{subsec:overview:static}

Next, we describe \lioHybrid, a static version of \gls{liostar}, 
where \gls{ifc} checks are statically verified by \gls{fstar}, 
instead of being tested at runtime, as in \lioDeian.

The crux of \lioHybrid is that the \gls{ifc} check performed 
by the |label| function is now lifted to a precondition of its |Ifc| effect~\citep{Maillard19}.
To call |label (v, l)|, the client needs to prove 
that the current label can flow to |l| as expressed by the |requires| statement |cur `canFlow` l| below. 
\begin{mcode}
  val label (v:'a) (l:'l) : Ifc (labeled 'a) 
    (requires fun cur -> cur `canFlow` l)
    (ensures fun li x lf -> labelOf x `eq` l /\ valueOf x == v /\ lf `eq` li) 

  val unlabel : (vl labeled 'a) : Ifc 'a
    (requires fun _ -> True)
    (ensures fun li x lf -> lf `eq` (li `join` labelOf vl) /\ x == valueOf vl)
\end{mcode}
The |Ifc| effect that allows the expression of such a requirement is defined in~\S~\ref{subsec:implementation:static}. 
To allow \gls{fstar} to discharge \gls{ifc} requirements expressed as 
pre-conditions, functions with an |Ifc| effect must stipulate a 
post-condition, by mean of an |ensures| clause, that exactly captures 
their behaviors, \ie the returned value and how the current label is modified. 
For instance, as specified above, |label| leaves the current label unchanged,  
while |unlabel| joins its initial label |li| with the label of its input |vl|.

\paragraph{Propagation of Proof Obligations}
The \gls{ifc} requirement of |label| is propagated to all its clients.
For example, the |checkLabeled| will not be verified unless 
we supply it with the |requires| below:
\begin{mcode}
  let checkLabeled (l:'l) (i:'a) (lv:labeled 'a): Ifc bool 
	(requires fun l0 -> l0 `canFlow` l)
	(ensures fun li x lf -> lf = li `join` l `join` labelOf lv) // definition as in 2.1

  let eqLabeled (v1 v2 :labeled 'a) : Ifc bool
	(requires fun _ -> True)
	(ensures fun li x lf -> lf = li `join` labelOf v1 `join` labelOf v2) // definition as in 2.1
\end{mcode}

To enable verification of its clients, 
|checkLabeled| also specifies the modifications it performs to the current label 
by an |ensures| post-condition. 
In turn, the verification of |checkLabeled|'s post-condition 
is only possible by the above guarantee of |eqLabeled|, 
which itself verifies by the guarantee of |unlabel|. 

\paragraph{Handing Dynamic Data with Path Sensitivity}
Of course, it is not always possible to statically discharge 
proof obligations, especially when they depend on dynamic data. 
For example, the function |dynCheck| below 
uses a function |getDynamicLabel|
that \textit{dynamically} returns a label |l| 
(\eg from the console, a database or a random input).
Then, it calls |checkLabeled| with that label. 
\begin{mcode}
  let dynCheck (i:'a) (lv:labeled 'a): Ifc bool = 
    let l  = getDynamicLabel() in // e.g., getDynamicLabel () = intToLabel(getRandom())
    let lc = getCurrent() in 
    if lc `canFlow` l 
      then checkLabeled l i lv 
      else ...
\end{mcode}
%
%
%
Since it is impossible to statically prove that 
the current label can flow to the dynamic |l|, 
then the call to |checkLabeled| must be guarded by the required runtime check.
Because verification in \gls{fstar} is path-sensitive, 
the call to |checkLabeled| will easily verify. 
The decision of what happens 
when the check fails is left to the user: 
if the user desires the code to be crash-free, then
they should ensure the else-branch be properly covered, 
an alternative that was not existent in \lioDeian. 

\paragraph{Runtime-check free C code}
Having paid the price of static \gls{ifc} check propagation and verification,
which in fact is highly aided by \gls{fstar}'s automation, 
we can enjoy C code without runtime checks. 
For example, the extracted C code for the |checkLabeled| function 
now simplifies to the below. 
\begin{mcode}
  bool checkLabeled__int32_t(labels l, int32_t i, labeled___int32_t lv) { 
	labeled___int32_t lv_ = { .data = i, .tag = l };
	return eqLabeled__int32_t(lv, lv_);
  }
\end{mcode}
Now, |checkLabeled| simply packs the inputs to a labeled value 
and calls |eqLabeled|. 
The extracted code, compared to \lioDeian, lacks all the runtime 
checks and exceptions derived from calls to |label|, leading 
to code amicable for unattended devices. 
But, code extracted from calls to the |unlabel| function remains unchanged. 
For instance, each call to |eqLabeled| is still updating the current label, 
which greatly slows down execution, especially in cases where it is actually unused.

\subsection{\lioStatic: ghosting the current label}
\label{subsec:overview:ghost}

The final variant of \gls{liostar} is \lioStatic which, 
in addition to removing the dynamic \gls{ifc} checks like \lioHybrid, 
 also removes all updates to the current label before runtime.
To do so, the current label is marked as computationally irrelevant, 
using \gls{fstar}'s |Ghost| module.  
In practice, this means that the current label is preserved at compile time, 
to verify  static \gls{ifc} proof obligations, 
but it is erased at runtime, thus the C extraced code is free from 
current state book-keeping. 

The implementation is presented in~\S~\ref{subsec:implementation:ghost}. 
and suffers from two inconveniences: 

\paragraph{Inconvenience I: Lifting from and to ghost values.}

The |Ifc| effect keeps track of a ghost state that preserves an |erased| label.
To pass this label to functions, \eg methods of the 
|lattice| type class that expect a label,
we need to use \gls{fstar}'s |reveal| function
that reveals the content of erased values within specifications. 
Dually, to turn labels into erased labels, \eg to compare them with the current label, 
we need to use \gls{fstar}'s |hide| function that erases values. 

For example, the specification of |checkLabeled| gets polluted with 
|reveal| and |hide|, as below:
\begin{mcode}
  let checkLabeled (l:'l) (i:'a) (lv:labeled 'a): Ifc bool // definition in 2.1
	(requires fun li -> reveal li `canFlow` l)
	(ensures fun li x lo -> lo = li `join` l `join` reveal (labelOf lv))
\end{mcode}
The semantics of the |requires| and |ensures| clauses remains the same, 
but now the argument label |l| exists at runtime, while 
the label arguments of the clauses are erased, 
thus conversions are required. 

\paragraph{Inconvenience II: Dynamic Checks require explicit book-keeping.}
\lioStatic turns really inconvenient when static \gls{ifc} checks involve dynamic data. 
For instance, consider that we want to replicate the function |dynCheck| 
from \lioHybrid, that calls |checkLabeled| with a label |l| obtained at runtime.
%
%
\begin{mcode}
  let dynCheck (i:'a) (lv:labeled 'a): Ifc bool = 
    let l  = getDynamicLabel () in 
    let lc = ??? // was getCurrent () in 2.2
    if lc `canFlow` l 
      then checkLabeled l i lv
      else ... 
\end{mcode}
In \lioHybrid, we used a runtime check to ensure that the current label can flow to |l|. 
This is not possible anymore, since the current label is erased at runtime. 
Now, what is a potential check we could perform to persuade \gls{fstar} 
that calling |checkLabeled| with the dynamic |l| is valid?

We can always construct a label |lc| so that the |checkLabeled| call verifies, 
by explicitly passing around the current label. 
This essentially means that the clients need to manually and correctly,
replicate all the current label accumulation that \lioHybrid was automatically doing. 
Of course, this is error prone and not encouraged, but suggests that 
handling dynamic data is possible in \lioStatic, though defeats its design. 
In short, unlike low-level applications, 
clients rely heavily on dynamic checks (as \acrshort{datastar} 
of~\S~\ref{sec:benchmarks}) will not benefit from \lioStatic. 

\paragraph{Current-label free C code}
Though strenuous at times, \lioStatic comes with the huge advantage 
that the extracted C code is completely free from label information.
For instance, below is the extracted C code 
from the functions |checkLabeled| and |eqLabeled|. 
\begin{mcode}
  bool checkLabeled__int32_t(int32_t i, int32_t lv) {
	int32_t lv_ = i;
	return eqLabeled__int32_t(lv, lv_);
  }

  bool eqLabeled_int32_t(int32_t v1, int32_t v2) {
	int32_t i1 = v1;
	int32_t i2 = v2;
	return i1 == i2;
  }
\end{mcode}
At runtime all labels are removed, thus labeled values 
are represented purely by their values. 
Thus, extraction to C erases not only all the current label book-keeping 
information, but also the data field selectors from labeled values. 
In short, the extracted code is light and amenable to low-level embedded systems, 
as supported by our benchmarks (\S~\ref{sec:benchmarks}). 
At the same time, \gls{ifc} checks have been verified to statically hold
by \gls{fstar}.

An additional benefit of \lioStatic
is that, due to its lack of stateful updates, it is amenable 
to prove metatheorems. 
Concretely, in~\S~\ref{section:metathery} we define a metaprogramming procedure that encodes 
a noninterference lemma on clients of \lioStatic. 
Intuitively, this procedure uses \gls{metastar} to derive 
1/ the ``low view'' of a \lioStatic function, 
\ie the view of an adversary with low privileges, and 
2/ a lemma that encodes that low view is preserved by evaluation, 
and thus, as in~\citet{Russo08}, encodes noninterference. 
We used this procedure on both the example 
functions |eqLabeled| and |checkLabeled|, as well as on two of our 
benchmarks (\S~\ref{sec:benchmarks}).
In all these cases, \gls{fstar} was able to automatically prove the noninterference, mechanically derived lemmata. 
	\section{Implementation of \liostar}
\label{sec:implementation}
	
This section presents the three versions of our \gls{ifc} library,
all of which express \gls{ifc} policies using  
a verified label type class described at~\S~\ref{subsec:implementation:labels}. 
Concretely, we define 

\noindent(\S~\ref{subsec:implementation:dynamic})
the dynamic \lioDeian, where \gls{ifc} policies are checked at runtime, 

\noindent(\S~\ref{subsec:implementation:static})
the static \lioHybrid, where \gls{ifc} policies are lifted to compile time proof obligations, and 

\noindent(\S~\ref{subsec:implementation:ghost})
the ghost \lioStatic, where the \gls{ifc} label tracking 
is ghosted, \ie totally erased at runtime. 

\subsection{Labels as a type class}
\label{subsec:implementation:labels}

\gls{ifc} policies are represented by a lattice~\citep{Denning:1976}.  
We encode the lattice of policies as an \gls{fstar} parametric type class  
that defines the lattice operations, constants, and properties. 
This encoding allows one to instantiate the type class to fit the policy 
of a domain-specific application, 
while ensuring all algebraic properties of the expected lattice operations 
are satisfied.

The ``can flow to'' operation (|`canFlow`|) defines the partial order relation between labels.  
The operations meet (|`meet`|) and join (|`join`|) respectively define the greatest lower bound 
and the least upper bound of the two labels in this lattice. 
Finally, bottom (|`bottom`|) is the minimum lattice element. 
As in~\cite{Parker:2019:LIF:3302515.3290388}, the type class is refined with the required lattice properties:
\begin{itemize}[leftmargin=*]
\item bottom is the smallest lattice value (|lawBot|), 
\item each label can flow to itself (|lawFlowReflexivity|),
\item if two labels can flow to each other they are flow-equivalent (|lawAntisymmetry|), 
\item if three labels can flow in chain then the minimum can flow to the maximum (|lawTransitivity|),
\item any label lower than a pair of labels can flow to the glb of these two labels (|lawMeet|), and 
\item the lub of a pair of labels can flow to any label accessible by this pair (|lawJoin|).
\end{itemize}		

\begin{mcode}
  class lattice a = { 
   `bottom`: a;
   `canFlow`: a -> a -> Tot bool;
   `meet`: a -> a -> Tot a;
   `join`: a -> a -> Tot a;

   lawBot: l:a -> Lemma (`bottom` `canFlow` l);   
   lawReflexivity : l:a -> Lemma (l `canFlow` l);
   lawAntisymmetry : x:a -> y:a -> Lemma ((x `canFlow` y /\ y `canFlow` x) ==> x `equals` y);
   lawTransitivity : x:a -> y:a -> z:a -> Lemma ((x `canFlow` y /\ y `canFlow` z) ==> x `canFlow` z);
   lawMeet : z:a -> x:a -> y:a
     -> Lemma (z `equals` (x `meet` y) ==>  z `canFlow` x /\ z `canFlow` y /\ (forall (l:a). l `canFlow` x /\ l `canFlow` y ==> l `canFlow` z));
   lawJoin : z:a -> x:a -> y:a
     -> Lemma (z `equals` (x `join` y) ==> x `canFlow` z /\ y `canFlow` z /\ (forall (l:a). x `canFlow` l /\ y `canFlow` l ==> z `canFlow` l));
  }
\end{mcode}
			
To generate an \gls{ifc} policy, one needs to instantiate the |lattice| type class with a concrete lattice.  
One such simple instance is, for example, the minimalistic classification of |Low| for public data, |Medium| for sensitive information, and |High| for private data:

\begin{mcode}
  type simple_lattice = | Low | Medium | High
\end{mcode}

The next step is to define the partial order |lcanFlow| using a comparison function |lt| 
between the three label values, and then the join and meet operations; as below:

\begin{mcode}
  let lt a b = match a, b with  
  	| Low, Medium -> true
  	| Medium, High -> true 
	| Low, High -> true 
	| _, _ -> false
  let lcanFlow a b = lt a b || a `eq` b
  let ljoin a b = if lt a b then b else a 
  let lmeet a b = if lt a b then a else b 
\end{mcode}
		
The above functions define an instance of the |lattice| type class. 
All the lattice's laws, \eg |Lemma (`bottom` `canFlow` l)|, are automatically proved 
by \gls{fstar} using the above function definitions.
Consequently, the instance does not need to provide any explicit proof terms, 
which are hence left as unit returning functions, \eg |$\lambda$ _ -> ()|.

\begin{mcode}
  instance SimpleLattice: lattice simple_lattice = {
	`canFlow` = lcanFlow;
	`meet` = lmeet; 
	`join` = ljoin;
	`bottom` = Low; 
	lawBot = ($\lambda$ _ -> ());
	lawReflexivity = ($\lambda$ _ -> ()); 
	lawAntisymmetry = ($\lambda$ _ _ -> ()); 
	lawTransitivity = ($\lambda$ _ _ _ -> ()); 
	lawMeet = ($\lambda$ _ _ _ -> ()); 
	lawJoin = ($\lambda$ _ _ _ -> ());
  }
\end{mcode}


To simplify away the |lattice| type class constraints,
our libraries behave as parametrized modules\footnote{Because parametrized modules \textit{à la} OCaml are not available in \gls{fstar}, we trick, instead, \gls{fstar} module system.} over a lattice type.
That is, in the rest we use the type |'l| to refer to some
|lattice| instance. 

\subsection{\lioDeian: \gls{ifc} with dynamic runtime checks}
\label{subsec:implementation:dynamic}

Next, we present \lioDeian, an \gls{ifc} library where 
security policies are dynamically checked with runtime checks. 
This library is basically \gls{lio}~\cite{giffin2012hails} 
where the Haskell |LIO| monad is encoded as an \gls{fstar} effect.  

Data protection in \lioDeian is implemented by wrapping data 
with a security label to form a labeled value. 
To protect labeled values from policy-invalid access, 
an abstract public data type, called |labeled|, hides the actual 
data structure of the |private| |type| definition, called |labeledTCB|. 
The type |labeledTCB| is private and belongs to the library's \gls{tcb}. 
As described in~\S~\ref{subsec:implementation:labels}, 
it uses |'l| as an abstract data type that instantiates the |lattice| type class. 
To access the tag field of labeled values we define the operation |valueOf|.
The data field is the goal of \gls{ifc} protection, thus cannot be accessed.

\begin{mcode}
  private type labeledTCB 'a = { data: 'a; tag : 'l; } 
  type labeled 'a = labeledTCB 'a
  let labelOf (vl:labeled 'a): 'l  = vl.tag
\end{mcode}

\paragraph{The IFC Effect}

To securely access the value of labeled data we define the |Ifc| effect  
that accumulates the highest label required to perform that access. 
The |Ifc| effect is a state effect that carries a |context| and is defined below.
\begin{mcode}
  type context = {cur: 'l}
  effect Ifc (a:Type) = IFC a (fun _ p -> forall r. p r)
\end{mcode}
The effect |Ifc| is parametric over the return type |a|
and has trivially true pre- and post-conditions. 
To keep the description simple, we present the |context| 
to only be  the current label, while our implementation also supports 
the LIO-style clearance optimization, defined as in~\cite{giffin2012hails}. 

The |Ifc| effect comes with two primitive operations that 
set and get the current label. 
%
\begin{mcode}
  private assume val setCurrent : 'l -> Ifc unit
  assume val getCurrent : unit -> Ifc 'l
\end{mcode}
The setter function is required to manipulate the context of the |Ifc| effect, 
but can be used to break policy enforcement by arbitrarily setting a low current label.
Thus, it is not exposed by our library, \ie is defined as |private|. 
Both of these functions are dropped during C extraction and are replaced with concrete C definitions.
%
\footnote{%
  Defining these setter and getter functions directly
  using the canonical \gls{lowstar} way of dealing
  with C memory would litter our code 
  with reasoning about the low-level memory model, 
  a problem that will be addressed by \gls{fstar}'s \href{https://github.com/FStarLang/FStar/wiki/Proposal:-Effect-Layering}{layered effect}, 
  currently under development.
}

To run an |Ifc| computation, one should extract his program (either in OCaml or in C); \gls{fstar} by itself is not intendeed at running code. 

\paragraph{Data Unlabelling}

To unlabel a labeled value |vl| the current label is raised to the label of |vl|, thus accumulating the privileges required to access 
|vl|. 
\begin{mcode}
  let unlabel (vl:labeled 'a) : Ifc 'a =
	raise (labelOf vl); 
	vl.data
\end{mcode}
Where the function |raise| sets the current label to the join of its argument 
and the old current label.  
\begin{mcode}
  let raise (l:'l) : Ifc unit = 
	let c = getCurrent () in  
	setCurrent (c `join` l)
\end{mcode}

\paragraph{Data Labelling}
The function |label v l|, labels the input value |v| with the input label |l|, 
when the current label can flow into |l|, otherwise it fails with a runtime error.
\begin{mcode}
  let label (v:'a) (l:'l) : Ifc (labeled 'a) = 
	let li = getCurrent ()  in  
	if li `canFlow` l 
	  then {data=v; tag=l}
	  else fail "invalid label";
\end{mcode}
Where the function |fail| is a wrapper around C's failing function, 
assumed to be in the |Ifc| effect: 
\begin{mcode}
  assume val fail : string -> Ifc 'a
\end{mcode}

\paragraph{Addressing the ``label creep problem''}
The ``label creep problem'' appears when 
an effectful computation, say |cmp|,  requires access to sensitive data 
and as a result raises the current label too high. 
As in~\cite{Buiras15}, to address this problem we define 
the |toLabeled| function that first computes the result of the sensitive computation 
|cmp|, then restores the current label to the original one, and finally 
returns the result of |cmp| labeled with the current label right after the 
|cmp| was computed. 

\begin{mcode}
  let toLabeled (cmp:b -> Ifc a) (params:b) : Ifc (labeled a) =
	let c0 = getCurrent () in     	(* save current label 						*)
	let v  = cmp params    in    	(* run cmp with the given parameters 	*)
	let c1 = getCurrent () in   	(* get new current label 					*)
	let vl = label v c1    in       (* result with the new current label 		*)
	setCurrent c0;              	(* restore the old current label, return 		*) 
	vl             
\end{mcode}
Since the result of |cmp| is labeled with the current label requires for 
|cmp|'s call and the current label is explicitly set to the original current labels, 
|toLabeled| addresses the label creep problem:
the result of |cmp| is properly protected, while the
current label remains unchanged. 

\paragraph{In short,}
\lioDeian is defined by naively implementing the original \gls{lio}
implementation to \gls{fstar}. 
Yet, even this naive porting provides a great benefit. 
Taking advantage of \gls{krml} one can export \lioDeian to high quality, 
low-level C code.

\subsection{\lioHybrid: \gls{ifc} with static proof obligations}
\label{subsec:implementation:static}
Next, we present \lioHybrid where the \gls{ifc} policy checks 
are turned into static proof obligations that are 
semi-automatically discharged by \gls{fstar}'s verification system. 
Intuitively, our goal is to lift the runtime check of |label|
into a static assumption that prevents failure. 
To do so, we need to 1/ refine the |Ifc| effect with a context propagation 
weakest precondition and 2/ refine all |Ifc| functions with 
descriptive assertions, so that static verification precisely
propagates the context information required to 
discharge |label|'s assumption.

\paragraph{The GTot effect to define specifications.}
In \gls{fstar}, specifications (namely \emph{type refinements}, \emph{ensures} and \emph{requires} clauses) 
belong to the |GTot| effect, which stands for |Ghost| and |Total|.  
Their purpose is to serve static verification and they are erased at compile-time.  
This clear boundary of what is and what is not used at runtime, 
allows us to define functions that provably will not be called at runtime. 
For example, below we define the |valueOf| accessor
\begin{mcode}
 let valueOf (vl:labeled 'a) : GTot 'a = vl.data
\end{mcode}
A call to |valueOf| at runtime will simply destroy the whole purpose of 
our library, since it unconditionally accesses protected data. 
Yet, wrapped within the |GTot| effect, we rest assured that |valueOf|
will only be used to define specifications and never leak the 
data of a labeled value at runtime.

\paragraph{Refinement of the Ifc effect}
The |Ifc| effect is now indexed with 
a pre and post-conditions defined below using 
\gls{fstar}'s Dijkstra Monads~\citep{dm4free}.

\begin{mcode}
  effect Ifc (a:Type) (pre:'l -> GTot Type0) (post:'l -> a -> 'l -> GTot Type0) =
	IFC a (fun (ci:context) (p:'a -> context -> GTot Type0) -> pre ci.cur /\
		    (forall (v:a) (co:context). 
			   (pre ci.cur /\ post ci.cur v co.cur) ==> p v co))
\end{mcode}
The |Ifc| effect definition is a standard state effect 
that consists of the return type |a| of the computation, 
the pre-condition |pre| on the the current label 
and the post-condition |post|.
The pre- and post-conditions return \gls{fstar}'s 
proposition type |Type0| within the |GTot| effect.   
Using \gls{fstar}'s Dijkstra Monads we define, in a standard way,
the weakest predicate transformer (WP) so that 
1/ the pre-condition is valid with the initial context label 
|pre (c0.cur)| and that 2/ for any returned value |v| and final context |co|
that satisfy the pre- and post-condition, 
then the post-condition |p v co| also holds. 

With this indexing, the |Ifc| functions 
can use precise |ensure| predicates to describe their behavior. 
For example, with refine the |getCurrent| and |setCurrent|
operations as follows.

\begin{mcode}
  assume val getCurrent : unit -> Ifc ('l)
	(ensures fun li x lo -> li `eq` lo /\  x `eq` lo)

  private assume val setCurrent : (l:'l) -> Ifc (unit)
	(ensures fun li _ lo -> lo `eq` l)
\end{mcode}
The function |getCurrent| ensures that 
the current label is preserved (\ie |li `eq` lo|) and returned 
(\ie |x `eq` lo|). 
The function |setCurrent l| ensures that the current label 
is set to |l| (\ie |lo `eq` l|).
\gls{fstar} also needs requires clauses, where 
when omitted the trivial |(requires fun _ -> True)| is implied. 


%

\paragraph{Data Unlabelling}
The definition of data unlabelling, \ie |unlabel|, 
and label raising, \ie |raise|, 
are exactly the same as in \lioDeian. 
But now, both functions come with specifications 
that precisely capture their behavior. 
\begin{mcode}
  let unlabel (vl:labeled 'a): Ifc ('a)
    (ensures fun li x lo -> lo `eq` li `join` (labelOf vl) /\ x == valueOf vl)

  let raise (l:'l) : Ifc unit
    (ensures fun li x lo -> lo `eq` li `join` l) 
\end{mcode}
The ensure clause of |unlabel| specifies that the output label 
is the join on the input and the label of the argument, 
while the returned value is the value of the argument. 
The verification of this definition is only possible when the |raise| function 
also comes with a precise ensure clause that, as above, states 
that the current label is joint with the function's argument. 

\paragraph{Data labelling} 
The crux of the \lioHybrid library is that the definition of 
|label| is now changed to be crash-freedom free. 
Now |label v l| simply labels the value |v| with |l|,
while its precondition ensures that no information is leaked. 
\begin{mcode}
  let label (v:'a) (l:'l) : Ifc (labeled 'a) 
	(requires fun li -> li `canFlow` l)
	(ensures fun li x lo -> (labelOf x) `eq` l /\ (valueOf x) == v /\ lo `eq` li) 
  = {data=v; tag=l}
\end{mcode}
\gls{ifc} is captured by the requirement that the current label 
should flow to the argument label |l|, while the |ensure| clause 
precisely captures the function's behavior of |label| 
to aid static verification of |label|'s clients.

\paragraph{Addressing the ``label creep problem''}
The |toLabeled| function that addressed the label creep problem 
is the same as before.
Yet, to pass static verification yet, its
type signature requires a slight modification.
Concretely, we use \gls{fstar}'s argument metaprogramming \$ key 
to define the type of |toLabeled| as follows: 

\begin{mcode}
  let toLabeled #a #b #pre #post
    ($\$$cmp:b->Ifc a (requires pre) (ensures post)) 
    (params:b)
    : Ifc (labeled a)
	(requires fun li -> pre li) 
	(ensures fun li x lo -> li `eq` lo /\ post li (valueOf x) (labelOf x)) 
\end{mcode}
The arguments |a| |b| |pre| |post| are implicit, thus 
marked with |#|.
The input computation |cmp| is infixed
with the \$ key that disable subtyping, inferring 
the implicit |pre| and |post| arguments based on 
|cmp|'s pre- and post-conditions. 
Then, the precondition is propagated as the |toLabeled|'s precondition 
to allow the call to |cmp| statically verify. 
The postcondition of |toLabeled| ensures that the current label is not modified. 
With this type signature, 
the definition of |toLabeled| is left the same as before and now 
the call of |cmp| is statically verified, since its precondition is propagated 
as a precondition to |toLabeled|. 

\paragraph{In short,} \lioHybrid has lifted the \gls{ifc} runtime check of |label|
to a static assumption. 
This assumption propagates to direct and indirect clients of 
|label|. 
To discharge these assumptions, all \lioHybrid functions come with  
precise postconditions. 
Then verification proceeds using the weakest preconditions of \gls{fstar}
as specified in the definition of the |Ifc| effect. 
This way, \lioHybrid clients are extracted to efficient C code 
that is runtime check free. 

\subsection{{\lioStatic :} label tracking becomes ghosted}
\label{subsec:implementation:ghost}
Finally, we present \lioStatic where all label tracking information is 
explicitly marked as computational irrelevant -- using \gls{fstar}'s ghost mechanism --  
and are thus removed from runtime. 

\paragraph{Computational Irreleval Values}
To encode computationally irrelevant values we use three functions below from \gls{fstar}'s |Ghost| module.  
Erased values are decorated with the erasable attribute |E|. The |reveal| function allows to access labeled values within the |GTot| effect, hence at compile-time only, whereas the |hide| function allows to erase them. 
\begin{mcode}
  module G
    type erased (a:Type) = | E of a
    val reveal : #a:Type -> erased a -> GTot a
    val hide   : #a:Type -> a -> Tot (erased a)
\end{mcode}

\paragraph{Ghosted Labeled Data}

The label field of a labeled value is now marked as erased: 
it is used to check policy enforcement at verification time 
and to make it explicitly unavailable at runtime. 
The definition and accessors of the labeled type are as follows: 
\begin{mcode}
  private type labeledTCB 'a = { data:'a;  tag:G.erased label; }
  type labeled 'a    = labeledTCB 'a (* public version *)
  let valueOf (vl:labeled 'a) : GTot ('a) = vl.data
  let labelOf (vl:labeled 'a) : (G.erased 'l) = vl.tag
\end{mcode}
The public version of the labeled type is, as before, 
a wrapper around |labeledTCB|, which now contains an erased label.
The value accessor remains unchanged, 
and can access the data at compile-time under the |GTot| effect. 
Importantly, the label accessor now returns an erased label, 
which pass be arbitrarily passed around, but 
can only be revealed in specifications, \ie under the |GTot| effect.

\paragraph{Ghosted \gls{ifc} Effect}
We further erase the current label field of the context.
\begin{mcode}
  type context = { cur:G.erased 'l;}
\end{mcode}
In our implementation, the context also contains 
a predicate over labels that we use to encode clearance-style optimizations.
Yet, for simplicity here we omit this field. 

We adjust the |Ifc| effect of \lioHybrid to account for the fact 
that now the current label is erased. 
Concretely, each access to the current label needs to also reveal 
its value, which is possible, since specifications live in the |GTot| effect. 
We declare |gc| as the function that first accesses the current label
and use it to define the |Ifc| effect as follows: 
  
\begin{mcode}
  let gc (c:context): GTot 'l = G.reveal c.cur
  effect Ifc (a:Type) (pre:'l -> GTot Type0) (post:'l -> a -> 'l -> GTot Type0) =
	IFC a (fun (ci:context) (p:'a -> context -> GTot Type0) -> pre (gc ci) /\
		    (forall (v:a) (co:context). 
			   (pre (gc ci) /\ post (gc ci) v (gc co)) ==> p v co))
\end{mcode}
Compared to the |Ifc| definition on \lioHybrid, 
we simply switched |c.cur| to |gc c|. 

The getter and setter functions of the ghost |Ifc| effect 
will also not be used at runtime. 
So, instead of wrapping C code (as in \lioHybrid)
now use the private |Ifc| functions |IFC?.get| and |IFC?.put| 
that are provided by \gls{fstar} as part of the effect definition. 
%
\begin{mcode}
  let getCurrent (_:unit) : Ifc (G.erased 'l) = 
	let cc = IFC?.get () in cc.cur
  private let setCurrent (l:G.erased 'l) : Ifc (unit) = 
    IFC?.put {cur = c}
\end{mcode}
The getter function |getCurrent| returns the erased current label, 
which can be revealed only at compile-time.  
The setter function |setCurrent| could still be used by clients to 
violate \gls{ifc} policies, so remains part of the libraries \gls{tcb}
and thus, is marked as private. 

The same holds for the |toLabeled| function, and in general each 
function that does not perform lattice operations 
(\eg |`canFlow`|, |`join`|) on erased labels.
When lattice operations are performed, like the functions
|raise|, |label|, and |unlabel| below, 
careful conversion is required between erased and actual labels.

\paragraph{Raising the Current Labels}
The function |raise| that is joining the current label on with 
its argument now takes as input an erased label. 
This edit is required, since in practice labels are generated 
after operations with the current label. 
Since the current label is not erased and cannot be revealed 
by clients without lifting the whole client as ghost, 
the label arguments to functions are turned into erased labels. 
Dually, it is always feasible to generate erased labels within the |Ifc| effect. 
That is because |hide|, that erases labels, 
is defined in the |Tot| effect, which is automatically lifted to |Ifc|. 

The definition and specification of |raise| is the following: 
\begin{mcode}
  let raise (l:G.erased 'l) : Ifc unit
    (ensures fun li x lo ->  lo `eq` li `join` G.reveal l) = 
	let li = getCurrent () in  
	setCurrent (G.hide (G.reveal li `join` G.reveal l))
\end{mcode}
The ensure clause simply reveals the argument label |l|, 
since it is in the |GTot| effect. 
To join the current labels |li| with the argument label |l|
both need to be revealed, producing the |GTot| effect. 
This effect though is encapsulated by the |hide| operation 
that turns the result on the join backed to erased.

\paragraph{Data Labelling and Unlabelling}
The definitions of data labelling and unlabelling
do not use any lattice specific operations, thus remain 
unchanged. 
Yet, their specifications require the addition of the reveal 
function each time erased labels are passed to lattice methods 
or compared. 
Concretely, the specifications of |unlabel| and |label| turn to the following:
\begin{mcode}
  let unlabel (vl:labeled 'a): Ifc ('a)
	(ensures fun li x lo -> lo `eq` li `join` G.reveal (labelOf vl) /\ x == valueOf vl)

  let label (v:'a) (l:G.erased 'l) : Ifc (labeled 'a) 
	(requires fun li -> li `canFlow` G.reveal l)
	(ensures fun li x lo -> G.reveal (labelOf x) `eq` G.reveal l /\ valueOf x == v /\ lo `eq` li) 
\end{mcode}
The only difference compared to the respective specifications of \lioHybrid
is that |G.reveal| is properly inserted on erased labels, before the lattice 
operations or equality is performed. 
This addition is an inconvenient, by type-directed process. 

\paragraph{Retrieving The Ghost Current Label}
The fact that in \lioStatic the run-time representation of the current label
is erased does not imply a loss of expressivity, since with great effort, 
one could emulate the current label at run time. 

In fact, \lioStatic is as expressive as \lioHybrid,
since \lioHybrid can be implemented as an external definition 
on the top of \lioStatic. 
Such a definition will take the shape of a state monad, 
carrying ---at runtime--- a label; 
it would provide wrappers for the |label|, |unlabel|, 
|raise| and |toLabeled| operations, acting as user defined proxies to 
\lioStatic's own operations. 
These definitions would be written so that the state monad would maintain 
a specific \gls{fstar} invariant: 
the runtime label stays equal to \lioStatic's erased label.

In practice, a \lioStatic user would selectively choose to pass around a 
runtime label that is provable to be equal to \lioStatic's erased label. 
Obviously, proving this label coherency property has the non-negligable human-time cost of writing proofs.

\paragraph{In short,} 
\lioStatic is using the ghost module of \gls{fstar} to 
mark the current label and the label inside labeled values 
as computation irrelevant, and thus erase it at runtime. 
The co-existence of erased and non erased labels 
makes writing specifications an inconvenient process, where 
reveal and hide annotations have to explicitly be provided. 
On return, 
as overviewed in~\S~\ref{sec:overview} and benchmarked in~\S~\ref{sec:benchmarks}
clients of \lioStatic translate to cleaner and more efficient, low-level C code. 
	\section{Meta-metatheory: Proofs of Non Interference}
\label{section:metathery}
In this section we make use of \gls{fstar}'s 
metaprogramming facilities (\acrshort{metastar}~\citep{metafstar})
to define a procedure, concretely the metaprogram |genNILemma|, that takes 
as input a toplevel name of a \lioStatic client, say |f|, 
and generates a lemma statement that |f| is noninterferent. 
We applied this procedure to two benchmarks from~\S~\ref{sec:benchmarks}
and the examples from~\S\ref{sec:overview}. 
In all these cases \gls{fstar} automatically proved 
the correctness of the mechanically derived, noninterference lemma.

\subsection{Statement of Non-interference}
\label{subsec:metatheory:statement}
Consider a \lioStatic function |f| that given an argument 
of type |'a| returns an |Ifc| computation with 
a precondition |pre|, postcondition |post|, and return value 
of type |b|.
\begin{mcode}
    val f : 'a -> Ifc b (requires pre) (ensures post)
\end{mcode}

We express noninterference of |f| 
using the low view preservation of~\citet{Russo08}.
That is, |f| is noninterferent when its low view 
is preserved by evaluation. 
This low view intuitively represents the view of an adversary 
with low privileges. 
More concretely, the low view on a level |l:'l| 
is defined by an \erase{l} function 
that forgets (\ie replaces by a ``hole'')
all the data that are protected by a label higher than |l|. 
The evaluation of the |Ifc| function |f| with an argument |x|
on the initial context |c| is denoted as \eval{f}{x}{c}.
With these notations, 
we express the noninterference lemma, 
\ie preservation of erasure by evaluation, as follows:

$$
\text{Lemma: }\ \text{fNI} \equiv
\forall \bind{\texttt{l}}{l}\ 
        \bind{\texttt{x}}{a}\
        \bind{\texttt{c}}{\texttt{context}\ \{ \texttt{pre c} \}}.\
\eraseapp{l}{\eval{f}{x}{c}} = \eraseapp{l}{\eval{(\eraseapp{l}{f})}{x}{c}}
$$

The above \fNI lemma states that the evaluation of |f| 
(\ie \eval{f}{x}{c} on the left hand side) 
and the evaluation of its erasure 
(\ie \eval{(\eraseapp{l}{f})}{x}{c} on the right hand side\footnote{
Note that the argument \texttt{x} does not need to be erased. 
Evaluation under erasure (\ie \eval{(\eraseapp{l}{f})}{x}{c})
ensures that each usage of the argument \texttt{x} on the erased \texttt{f} 
will be actually erased.})
cannot be distinguished after erasure, 
for any 
erasure level |l|,
input |x|,
and initial context |c| that satisfies |f|'s precondition.  

Our goal is to define a metaprogram that takes as input 
the binder of a function |`f| and encodes the noninterference lemma for |f|. 
To do so, we need to encode evaluation and erasure in \gls{fstar}. 

\paragraph{Encoding of evaluation}
In \gls{fstar} evaluation of effectful computations
is encoded by \citet{filinski94}'s reification
that changes effectfull calls 
from implicit monadic style to explicit passing style.
That is, 
the evaluation \eval{f}{x}{c} in the noninterference lemma \fNI
is encoded in \gls{fstar}
as |reify (f x) c|.

\paragraph{Encoding of erasure}
In \fNI erasure is used to erase both 
1/ the results produced after evaluation and 
2/ the function |f| itself. 
For the first case we define 
the |eraseCtx| function below.

\noindent
\begin{minipage}{0.65\textwidth}
\begin{mcode}
  let eraseCtx [|hasEraser 'a|] (l:'l) (x:'a,c:context) : 'a 
    if c.curr `canFlow` l then erase x else `hole`
\end{mcode}
\end{minipage}%
\begin{minipage}{0.35\textwidth}
\begin{mcode}
  class hasEraser 'a
    { erase :: 'l -> 'a -> 'a }
\end{mcode}
\end{minipage} 

\noindent
The function |eraseCtx| on the left,
takes as input an erasure level |l|
and the reification pair |(x,c)|. 
If the label of the input context can flow to |l| 
(\ie |c.curr `canFlow` l|), then the value |x| is 
erased, using the erasure method |erase| 
we define on~\S~\ref{subsec:metatheory:valueerasure}.
Otherwise, a `hole' (also defined on~\S~\ref{subsec:metatheory:valueerasure})
is returned. 

When erasure is used on the function |f| itself, 
\ie the appearance $\eraseapp{l}{f}$ in the \fNI lemma, 
it returns a different function, named |f_erased l|, 
that our metaprogramming procedure systematically generates. 
Intuitively, the top level function |f| with type |t|, 
generates a new top level function 
named |f_erased| with type |'l -> t|, \ie the erasure level is explicitly
passed, where all sensitive data protected with labels 
above |l| are erased. In~\S~\ref{subsec:metatheory:functionerasure}
we describe the generation of erased functions. 

\paragraph{NI Lemma Generation}
To generate the noninterference lemma
we call the metafunction |genNILemma|.
\begin{mcode}
  val genNILemma : Term -> Tac unit 
\end{mcode}
|genNILemma| takes as input the binder representation of a top level 
\lioStatic function, \eg |`f|, and generates a top level declaration 
that defines 
the noninterference lemma 
of |f|, named as |f_NI|. 
|genNILemma| takes as input a |Term|, that is
the meta-representation of an \gls{fstar}'s AST defined in \gls{metastar}, 
and has the |Tac| effect that allows 
inspection of the representation of values as well as function definitions, 
similar to TemplateHaskell's |Q| monad~\cite{templateHaskell}. 
It returns a |unit| value, since the top level |f_NI| lemma generation 
is an effect. 

The function |genNILemma| is 
a meta-program that generates one or multiple top level definitions,
executed by the means of the instruction |splice|. 
For example, |splice[](genNILemma `f)| generates the non-interference lemma 
below. The |[]| part of |splice|, required for scoping, 
is in the rest left empty for simplicity. 
\begin{mcode}
  %splice[](genNILemma `f) \\ generates the below lemma 
  let f_NI (l:'l) (x:a) (c:ctx{pre c})
   : Lemma (eraseCtx l (reify (f x) c) == eraseCtx l (reify (f_erased l x) c))
   = ()
\end{mcode}

The lemma generation metaprogram |genNILemma| 
supports functions with many arguments.
Concretely, it inspects the type of the input binder, 
and if it has |n| arguments,
it creates the lemma arguments |x1|..|xn| with the proper types 
and uses them to call the original and erased functions. 

For example, for the definitions of |eqLabeled| and |checkLabeled|
of the overview (\S~\ref{sec:overview}), the respective generated lemmata 
are shown below.
\begin{mcode}
  %spice[](genNILemma `eqLabeled_NI `checkLabeled) \\ generates the below lemmata 
  let eqLabeled_NI (l:'l) (x1:labeled 'a) (x2:labeled 'a) (c:context)
   : Lemma (eraseCtx l (reify (eqLabeled x1 x2) c) == 
            eraseCtx l (reify (eqLabeled_erased l x1 x2) c))
   = ()

  let checkLabeled_NI (l:'l) (x1:'l) (x2:'a) (x3:labeled 'a) (c:context {l.curr `canFlow` x1})
   : Lemma (eraseCtx l (reify (checkLabeled x1 x2 x3) c) == 
            eraseCtx l (reify (checkLabeled_erased l x1 x2 x3) c))
   = ()
\end{mcode}
The erased functions |checkLabeled_erased| and |eqLabeled_erased|
are defined at top level by |genNILemma| and 
are presented in~\S~\ref{subsec:metatheory:functionerasure}.

\gls{fstar} was able to automatically prove both the above lemmata, 
\ie with the trivial unit body definition, 
thus showing that neither of these functions interferes. 
Further, using |genNILemma| we proved two out of our three 
benchmarks (\acrshort{busstar} and \acrshort{mmustar})
are noninterferent, while the third (\acrshort{datastar})
imposes technical limitations, we discuss in~\S~\ref{subsec:metatheory:limitations}.

\subsection{Erasure of Values}
\label{subsec:metatheory:valueerasure}
Here we describe the definition of the |erase l x|
methods from~\S~\ref{subsec:metatheory:statement}
that is erasing, \ie replacing by a ``hole'', 
all data inside the value |x| that are protected by a label higher than |x|.

The hole is encoded in \gls{fstar} 
as |`hole`|, \ie an axiomatized polymorphic value defined below. 
\begin{mcode}
  assume val `hole`: ('a: Type) -> 'a
\end{mcode}
Thus |`hole`| has no content and can used to replace 
sensitive data of any type. 
 
\paragraph{Erasing a labeled value.}
To erase a labeled value, we define the function |eraseLabeled| below 
\begin{mcode}
  let eraseLabeled [|hasEraser 'a|] (l:'l) (x:labeled 'a) : labeled 'a =
    if x.tag  `canFlow` l 
      then { data=erase l (x.val); tag=x.tag } 
      else { data=`hole`; tag=x.tag }
\end{mcode}
Erasure of the labeled value |x| 
checks if the |tag| of |x| can flow to the erasure level |l|. 
If it can, it recursively erases the data, 
otherwise it replaces the |x|'s data with a |`hole`|. 
At each case, 
the |tag| field remains untouched. 
There are two things to note here:
First, our implementation forcefully violates 
the |labeled| privacy, specified in the \lioStatic implementation. 
Second, all the erased labeled (\ie in the tag field) are revealed.
Both of these are implemented using a trick on \gls{fstar} module system
to emulate lacking OCaml-like module functors. 

We use the |eraseLabeled| definition 
to define the |hasEraser| instance on the |labeled| type. 
\begin{mcode}
  let labeledHasEraser [|hasEraser 'a|]: hasEraser (labeled 'a) = {
    erase l x = eraseLabeled l x 
  }
\end{mcode}

For the rest types the |hasEraser| instance definition acts as a 
homomorphism.
For \textit{primitive types} (like $\mathbb B$, $\mathbb Z$ or $\mathit{unit}$), 
we defined the |hasEraser| instance to be the identity. 
For \textit{inductive data types} were defined a metaprogram 
that using generic programming techniques automatically defines 
the |hasEraser| instances.
For example, for lists, our metaprogram follows the list structure 
to mechanically generate the |eraseList| below. 
\begin{mcode}
  let rec eraseList [|hasEraser 'a|] (l:'l) (x:list 'a) : list 'a =
    match x with
      | Cons hd tl -> Cons (erase l hd) (eraseList l tl)
      | Nil -> Nil
\end{mcode}
The |eraseList| is used, like in |eraseLabel|, to define the 
list |hasEraser| instance.

In short, our metaprogram mechanically generates 
|hasEraser| instances for inductive types, while 
generation of instances for other types (\eg arrow types or higher kinded types),
when required, is left to the user. 

\subsection{Erasure of Functions}
\label{subsec:metatheory:functionerasure}
Function erasure is a metaprogram, 
|eraseF :: Term -> Tac unit| that given a
\lioStatic top level binder, say |`f|,  generates 
the top level definition |f_erased| as follows. 
The first argument of |f_erased| is a label |(lErase : 'l)|. 
We define erased function parametric on erasure level 
to avoid definition of multiple functions for different erasure levels. 
Then, the specifications are left unmodified, while the body definition 
of |f_erased| follows the AST of |f| where 
1/ each labeled subterm  is erased and 
2/ each function binder is replaced by its erased version, 
as explained below. 

\paragraph{1. Erasure of Labeled Subterms}
If |e| is a labeled subterm in the function definition, 
it is replaced by |eraseLabeled lErase e|, 
where |lErase| is the erasure level argument introduced to |f_erased|. 
To check if |e| is a labeled term, |eraseF| defined in the |Tac| effect, 
simply type checks the expression 
|eraseLabeled lErase e|. 
Thus, for each subterm |e|, if |eraseLabeled lErase e| type checks
it always replaces |e|, leading to some benign extra checks.  
Of course, the subterms |e| are open, \ie have unbounded variables, 
thus during AST traversal we keep an environment of 
the introduced variables and use it to type check subterms. 

\paragraph{2. Erasure of Function Binders}
If |g e1 ... en| is a subterm of the original function, 
with |g| being a function symbol, then |g| also needs to be erased. 
By default, 
we replace the function call with 
|g_erased lErase e1 ... en| and recursively call
|eraseF| on |`g| to define the erased declaration of |g|, 
if it is not already defined. 
We keep a list of function binders that do not need 
to be erased and calls to such functions remain untouched. 
First, we do not erase the functions defined in the \lioStatic
since their erasure is provably an identity, 
as captured by their specifications.
Second, we do not erase functions imported
from \gls{fstar} standard libraries, \eg map, |=|, |+|, \dots

\paragraph{Axiomatization of Contamination}
The decision not to erase primitive \gls{fstar} functions 
is made to avoid code expansion, and highlights a problem 
we call \textit{contamination}.
Contamination captures the |`hole`| propagation from 
arguments to results, \eg |`hole` = 42| should be equal to |`hole`|.
In general, if a function |g| \textit{consumes} it's $i$th argument,
then the call of |g| with a |`hole`| on the $i$th position, 
should be equal to |`hole`|. 
Our implementation axiomatizes contamination for primitive functions, 
\eg below we provide the contamination axioms for |=| and |+| 
on their first argument. 
\begin{mcode}
  let contaminationEq1   n : Lemma (`hole` = n == `hole`) = admit ()
  let contaminationPlus1 n : Lemma (`hole` + n == `hole`) = admit ()
\end{mcode}
Note that contamination axioms on the second argument are not required, 
since the SMT solver will derive them by commutativity.
Contamination axioms are not required for most functions from the standard library, 
\eg |map| and |fold| that can be normalized.
To aid contamination axiom generation and limit errors
we mechanised axiom generation by the implementation of a metaprogram
that given the function name and contamination position 
derives the proper contamination axiom. 

\paragraph{Examples of Function Erasure}
To illustrate the function erasure process, 
reconsider the |checkLabeled| and |eqLabeled| functions
from the overview (\S~\ref{section:overview:eqlabeled}). 
Their erased version produced by calling |eraseF| via 
|splice| are presented below. 
\begin{mcode}
  %splice[add_erased, checkLabeled_erased] (eraseF (lookup (`checkLabeled)))

  let eqLabeled_erased (lErase: 'l) (v1 v2:labeled 'a)
    : Ifc bool = 
    let i1 = unlabel (eraseLabeled lErase v1) in
    let i2 = unlabel (eraseLabeled lErase v2) in
    i1 = i2
        
  let checkLabeled_erased (lErase: 'l) (l:'l) (i:'a) (lv:labeled 'a): Ifc bool 
    (requires fun li -> reveal li `canFlow` l)
    (ensures fun li x lo -> lo = li `join` l `join` reveal (labelOf lv))
    let lv' = eraseLabeled lErase (label i l) in
    eqLabeled_erased lErase (eraseLabeled lErase lv) (eraseLabeled lErase lv')
  \end{mcode}
From the above examples we note that 
the functions specifications remain unchanged, though now |reveal|
and |hide| are now identities, as discussed in~\S~\ref{subsec:metatheory:valueerasure}.
The erasure level argument |lErase| is passed around
to prevent multiple definitions of erased functions on different 
erasure levels.
The functions |label|, |unlabel| are not erased because the belong in \lioStatic
and |=| is not erased because it is a primitive \gls{fstar} function 
(it is though axiomatized for contamination).
The function |eqLabeled| is erased. 
If fact, in the above |splice|
the original call of |checkLabeled| to |eqLabeled| triggered 
the declaration of |eqLabeled_erased|. 
Finally, we observe a redundancy on |eraseLabeled| wraps.
The binder |lv'| is wrapped both at its definition and call site.
Such redundancies can be syntactically detected and removed, 
thought they do not affect our goal that is static proof of noninterference.

\subsection{Discussion \& Limitations}
\label{subsec:metatheory:limitations}

The implementation of the metaprogramming procedure 
for noninterference lemma extraction |genNILemma| is about 800 Lines of Code 
(without spaces, with comments). We abstracted the \lioStatic 
dependencies of the implementation in such a way that
|genNILemma| can be easily adapted by other \gls{ifc} libraries. 
Due to continuing modifications of the API of \gls{metastar},
our implementation is build against a specific \gls{fstar} built\footnote{
  We use a patched version of the \gls{fstar} commit \texttt{0362a90a83bea851fa5e720637f1cb9d3dfe97bc}, for more detail see the Nix expression in the implementation sources.
}.

Our approach has three main limitations. 

\paragraph{Immaturity of \gls{metastar}}
Our implementation is a heavy client of \gls{metastar} which is quite immature. 
Published in~\citeyear{metafstar}, \gls{metastar}
has still various limitations, most importantly the  
inability of full AST inspection 
(\eg arbitrary effect weakest preconditions cannot get inspected).
More, \gls{metastar} is rapidly changing, which makes 
development and use of our library inconvenient.
Yet, exactly due to these changes, we believe 
that is the near future all the current limitation will get addressed.

\paragraph{Reifiable Requirement}
The main limitation of |genNILemma|
is that, to state noninterference of an |Ifc| function |f|
we use reification of |f|, 
thus |f| should live in an effect that is reifiable 
to a total computation. 
For example, potentially diverging expressions cannot 
be handled |genNILemma|, which is expected
since most noninterference proofs are 
termination sensitive~\cite{stefan:2017:flexible,Parker:2019:LIF:3302515.3290388}.
This limitation is the reason why we did not apply the \acrshort{datastar}
benchmark (from~\S~\ref{sec:benchmarks}) to |genNILemma|. 
Our benchmark needs access to a database, thus 
has an effect that is not directly reifiable to total.  

There is a way to address this limitation, which is 
to \textit{assume} the nonreifiable effect using a specification. 
That is, to give static semantics by means of a dependent-type, 
but no dynamic semantics, \ie no implementation.
In such a case, one is able to conduct proofs, 
but unable to normalize programs (as some primitives have no implementation), 
which is necessary to properly conduct the proofs 
on generated theorems.
In the near future, we plan to follow this approach and apply |genNILemma| 
to our \acrshort{datastar} benchmark. 

\paragraph{Specificity}
The generated noninterference proofs are very specific. 
Each proof is developed for exactly one client of \lioStatic. 
But, our metaprogram is defined for every \lioStatic client 
constrained by the above limitations. 

Our decision to specifically target clients of \lioStatic
instead of proving the correctness of the \lioStatic library 
stems from the fact that our goal is to verify real executable code. 
This is difficult, since real code comes with various verification 
unfriendly constructs. 
Thus, instead of following the route 
of~\cite{stefan:2017:flexible,Parker:2019:LIF:3302515.3290388}, 
that is to design a model of LIO and prove it correct, 
we restricted ourselves to per-client proof construction. 
This approach comes with another huge benefit. 
Both library noninterference proofs 
of~\cite{stefan:2017:flexible,Parker:2019:LIF:3302515.3290388}
impose limitation on the programs that do not interfere, 
concretely, those programs should be terminating 
and ``safe'' which intuitively means that 
they should not access the library's \gls{tcb}. 
In our proof such assumptions do not exist. 
Instead, if |genNILemma| is called on programs 
that violate the safety assumption,
\gls{fstar} would not be able  
to prove correct the derived noninterference lemma.

In short, our metaprogramming approach
targets clients of the real \lioStatic library 
instead of proving the correctness of a verification 
friendly model of the library. 
The noninterference lemma is mechanically derived 
per client, by a simple call to |genNILemma| 
and the proof is automated by \gls{fstar}.
	\section{Benchmarks}\label{sec:benchmarks}
	
In this section 
we present three benchmarks: 
\gls{datastar} in \S~\ref{bench:datastar}, 
\gls{busstar} in \S~\ref{bench:busstar}, 
and \gls{mmustar} in \S~\ref{bench:mmustar}
and explore how their implementation on the three versions 
of our library 
(\lioDeian, \lioHybrid, and \lioStatic)
affects the size of the development 
\gls{fstar} code, the size of the extracted C code, 
as well as the runtime performance of the benchmarks. 

	
\subsection{The \gls{datastar} case study}
\label{bench:datastar}
\label{subsec:bencmarks:db}
			
Our first case study is \gls{datastar}, which is inspired by 
the $\lambda$Chair conference management system of~\cite{stefan:2017:flexible}. 
The goal of \gls{datastar} is to perform the transactions of a conference selection process: 
submission, assignment, review, selection, results, while maintaining anonymity 
of reviewers and respecting conflicts of interest.
We ported $\lambda$Chair examples to all three versions of \gls{liostar}. 

The example below is using \lioStatic to implement the scenario where 
the user |Charles| reviews all the papers submitted by the user |Mary|. 
Concretely, the call |fetchPapers_for Mary| brings all the |labeled|
papers of |Mary| and the call |map add_review_from_Charles|
adds a review from |Charles| for each of the labeled papers. 
\begin{mcode}
  let example (_:unit) : Ifc unit
	(requires fun _ -> True)
	(ensures fun li _ lf -> li == lf) = 
	let l = fetchPapers_for Mary in	 (* get all Mary's labeled papers             *)
	(* the current label does not grow because the papers are not opened          *)
	map add_review_from_Charles l    (* Charles will add his review to each paper *) 
	(* the map executes toLabeled for each paper so the context is still preserve *)
\end{mcode}
The |ensures| clause above requires that the current label is preserved. 
Since |fetchPapers_for| returns labeled papers it does not 
raise the context, but for reviewing higher privileges are required. 
Locally raising the current label can only be achieved by the |toLabeled| function.

When we use \lioStatic, where labels of |labeled| values are ghosted
we face a problem. We are required to locally raise the current label
using the dynamic labels that protect |Mary|'s papers, which is impossible. 
To address this problem, in this \gls{datastar} benchmark, we boxed |labeled|
values with a label, that exists at runtime and \textit{soundly} approximates 
the ghost label of the |labeled| value. 
We define the |box| type as follows: 

\noindent
\begin{mcode}
  type box a = {
    data: labeled 'a; 
    tag: (l: 'l{gc (labelOf data) `canFlow` l}) 
  }

  let unbox (bv:box 'a) : Ifc 'a
  (ensures fun li x lf ->  (gc lf) `eq` ((gc li) `join` bv.tag) /\ x == valueOf (bv.data)) = 
    unlabel bv.data
\end{mcode}
The refinement type in the |tag| of the |box|
ensures that the |box|'s tag can be safely used at runtime 
when the protection label is required. 
In the \lioStatic implementation of \gls{datastar} 
we store |box| values in the database -- to be able to access the labels at runtime --  while 
the \lioDeian and \lioHybrid implementations store |labeled|
values. 

In~Table~\ref{table:benchmark_size} we observe 
that the lines of \gls{fstar} program code (LoP) of the implementation 
increased from 166 in \lioDeian (which is very similar to the original \gls{lio} implementation)
to 265 in \lioHybrid and to 269 in \lioStatic. 
This is a big increase on  \gls{lop} which depicts that 
the verification effort required was strenuous. 
The reason for that is that this specific example was developed 
to present the expressiveness power of \gls{lio} and it is very heavy on 
dynamic \gls{ifc} checks. 
In~Table~\ref{table:benchmark_speed} we observe that this verification 
effort does not pay off at runtime speedups which are only 9.1\%
and 12.6\% for \lioHybrid and \lioStatic, respectively. 

From this benchmark we conclude
that all the libraries are equally expressive, 
but clients with heavy runtime checks are advised to use the dynamic 
\lioDeian version. 

\subsection{The \gls{busstar} case study}\label{bench:busstar}

\begin{wrapfigure}{r}{.55\textwidth}
	\centering\includegraphics[width=.45
	\columnwidth]{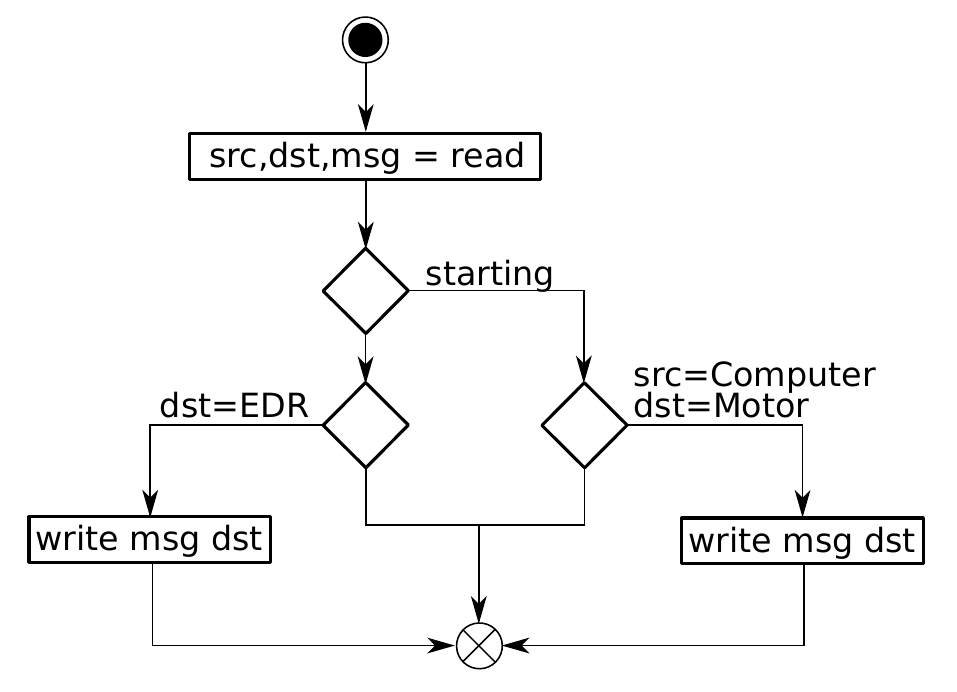}
	\caption{Flowchart of {\tt readBUS}}
	\label{fig:ifc_bus_receive}		
\end{wrapfigure}

Our second benchmark is an \gls{busstar} \gls{api}
that uses \gls{ifc} to ensure component separation. 
Figure~\ref{fig:ifc_bus_receive} gives flowchart of the \gls{busstar}'s algorithm
which is triggered any time a packet is available for transport. 

We enforce component separation using \gls{liostar} as follows. 
For each |actor|, \ie a system component, we define a security label (\ie |type component = 'l|). 
The current label of the |Ifc| effect keeps track of the accumulated BUS label. 
Each time an |actor| reads from the BUS,  
the current label is raised with |actor|. 
Dually, each time an |actor| writes to the BUS, 
it is required that the current label can flow to the |actor|. 
This behavior is similar to \gls{lio}'s |unlabel| and |label| operations. 
Thus, to enforce component separation we simply 
wrap the BUS read and write primitives to the |Ifc| effect 
via |unlabel| and |label| operations. 
Below we provide the wrappers as implemented in \lioStatic. 
\begin{mcode}
  let writeBUS (actor:component) (data:byte) : Ifc unit
	(requires fun li -> li `canFlow` actor) (ensures fun li x lo -> li `eq` lo) =
  	write actor (label data (G.hide actor))

  let readBUS (actor:component): Ifc byte 
	(ensures fun li x lo -> lo `eq` li `join` actor) =
    unlabel (read actor)
\end{mcode}

%

\begin{figure}[ht]
	\centering\includegraphics[width=.8
	\columnwidth]{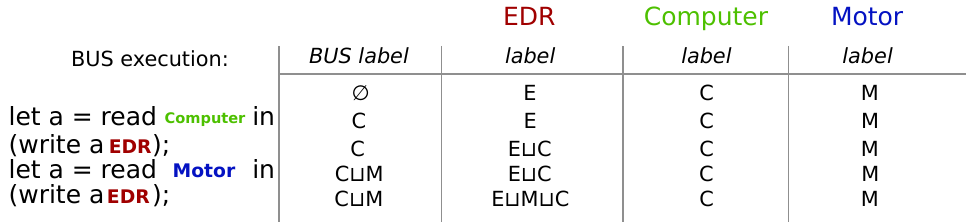}
	\caption{Event Data Record Example using in \gls{busstar}}
	\label{fig:ifc_bus_sequence}
\end{figure}

\begin{wrapfigure}{r}{.4\textwidth}
	\centering
	\includegraphics[width=.4\textwidth]
	{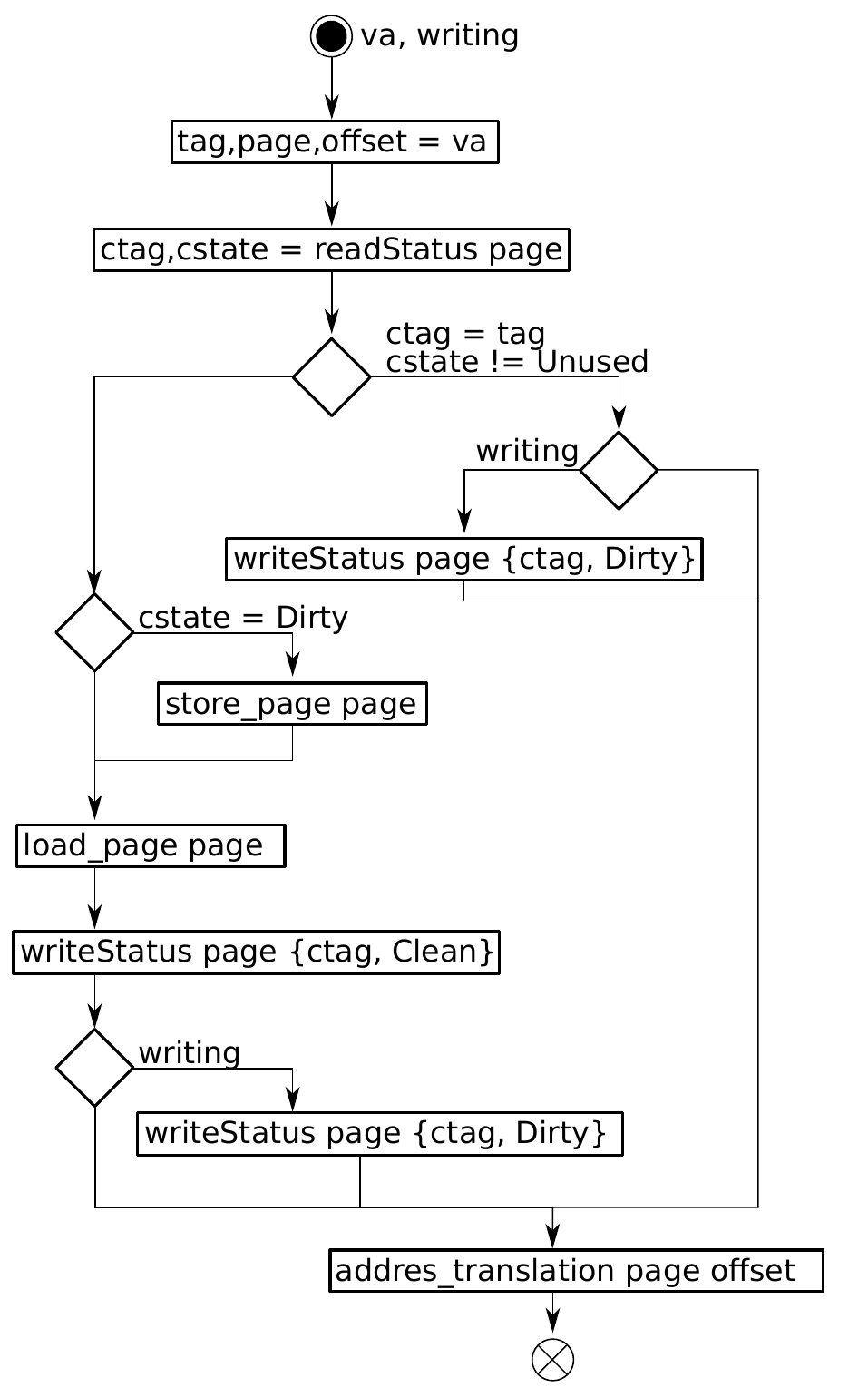}
	\captionof{figure}{Specification of \gls{mmustar}'s page swap}
	\label{fig:ifc_mmu_algo}
\end{wrapfigure}

Figure~\ref{fig:ifc_bus_sequence} presents an \gls{edr}
example, where \gls{busstar} controls the transport of data from 
the computer, the motor controller and the \gls{edr}.  
Using \gls{busstar} to implement an \gls{edr} requires two \gls{ifc} policies. 
First, when the system starts, the computer is allowed to communicate with the engine. 
Second, the computer and the engine cannot exchange any information but can only communicate 
with the \gls{edr} to register event data.

We implemented this \gls{edr} example in \lioDeian, \lioHybrid and \lioStatic. 
In~Table~\ref{table:benchmark_size} we observe that our implementation is 94 LoP
in all versions, since the client did not require any static specifications.
The lines of extracted C code  \gls{loc} greatly reduced from 
100 \gls{loc} in \lioDeian, 
to 92  \gls{loc} in \lioHybrid, and
to 68  \gls{loc} in \lioStatic. 
This reduction expresses that the few runtime checks imposed by our example
were trivially verified, rendering runtime label book tracking useless. 
In~Table~\ref{table:benchmark_speed} we observe that the running time
of the example is in accordance with the extracted C code size and  
gives 21.4\% speedup for \lioHybrid and 54.6\% for \lioStatic.

This benchmark illustrates that 
for applications that are not heavy on \gls{ifc} checks
and do not depend on dynamic labels (in contrast to~\S~\ref{subsec:bencmarks:db}), 
moving from \lioDeian to \lioStatic gives significant runtime speedup 
with trivial verification cost effort.

\subsection{The \gls{mmustar} case study}\label{bench:mmustar}

Our third benchmark is \gls{mmustar} that uses \gls{ifc} policies
to isolate application memory resources. 
Our implementation is inspired by \cite{choudhuri2005software}, 
that introduces the principle of a software \gls{mmu} for embedded 
devices that lack hardware virtualization.

\gls{mmustar} allows us to use virtualization to isolate tasks but also to permit data transfer between tasks memory.  When writing to a memory location, a program must craft a virtual address with a |tag|: the target task's id, a |page_index|: the page to write, and an |offset_index|. This allows any tasks to read and write on any memory location depending on the policy.  

In order to perform the transaction to physical memory, the \gls{mmustar} translates the virtual address to a physical one using a private function called |translation|. This function takes two arguments: a virtual address and a boolean that indicates the type of translation (read or write).  
Figure \ref{fig:ifc_mmu_algo}, illustrate the translation algorithm.

Our example, has two tasks: 
|task1| which writes two bytes in the memory of the second task and 
|task2| which reads these two bytes and performs an addition. 
We use \gls{liostar} to show that |task2| reads only its own memory, 
in a way similar to \gls{busstar} presented before (\S~\ref{bench:busstar}). 

In this example, as in \gls{busstar}, we observe C code reduction 
and runtime speedup when \gls{ifc} checks are statically proved. 
As~Table~\ref{table:benchmark_size} presents our implementation is 124 LoP
in all versions, while  \gls{loc} greatly reduced from 
192 \gls{loc} in \lioDeian, 
to 176 \gls{loc} in \lioHybrid, and
to 155 \gls{loc} in \lioStatic. 
As with \gls{busstar}, the C code reduction 
leads to runtime speedups of 23\% for \lioHybrid and 47.7\% for \lioStatic
shown in ~Table~\ref{table:benchmark_speed}.
Thus, again we get important runtime speedup with low verification effort. 




\subsection{Evaluation}
\label{subsec:bencmarks:eval}

\begin{wrapfigure}{r}{.58\textwidth}
	\centering
	\begin{tikzpicture}[scale=.625]
	\begin{axis}[
	ybar, bar width=8pt,ymin=0,
	xmin=0.5,xmax=3.5, xtick=data,
	xticklabels from table={\locdata}{paradigm},
	xticklabel style={rotate=45,anchor=north east,inner sep=0mm},
	ylabel={\Large \gls{loc} (lines)}, ylabel near ticks]
	\addplot table [x expr=\coordindex+1,y=lio] {\locdata};
	\addlegendentry{\lioDeian};
	\addplot table [x expr=\coordindex+1,y=lioC] {\locdata};
	\addlegendentry{\lioHybrid};
	\addplot table [x expr=\coordindex+1,y=lioS] {\locdata};
	\addlegendentry{\lioStatic};
	\end{axis}
	\end{tikzpicture}
	\begin{tabular}{r|rr|rr|rr|}
		& \multicolumn{2}{c|}{\lioDeian} & \multicolumn{2}{c|}{\lioHybrid} & \multicolumn{2}{c|}{\lioStatic} \\ \cline{2-7} 
		&  \gls{lop}     &  \gls{loc}    &  \gls{lop}     & \gls{loc}     &  \gls{lop}    & \gls{loc}     \\ \hline
		\gls{datastar} (\S~\ref{bench:datastar})	&  166        & 456            &    265      &  426           & 269         &    414      \\
		\gls{busstar} (\S~\ref{bench:busstar}) & 94 & 100 & 94 & 92 & 94 & 68 \\
		\gls{mmustar} (\S~\ref{bench:mmustar}) & 124 & 192 & 124 & 176 & 124 & 155 \\ \hline
		Total & 384 & 748 & 483 & 694 & 487 & 637 \\ \hline
	\end{tabular}
	\caption{Size measurement of all presented uses cases.}
	\label{table:benchmark_size}
\end{wrapfigure}
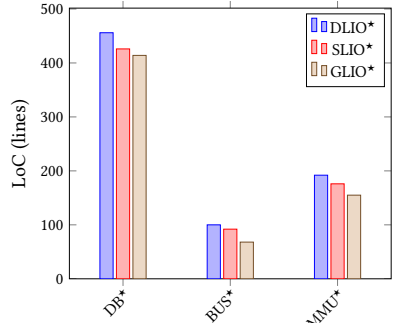

Finally we summarize the implementation platform 
and evaluation of our libraries. 

Figure~\ref{table:software_used} presents the hardware and software 
for the development and evaluation of our libraries. 
We used the \gls{gcc} compiler because our work is 
targeting micro-controller like Amtel which are not supported by 
the verified compiler CompCert.
For our benchmarks, we explicitly disable any compiler optimizations.

\begin{figure}[ht]
	\centering
	\begin{tikzpicture}[scale=.625]
	\begin{axis}[
	ybar, bar width=8pt,ymin=0,
	xmin=0.5,xmax=3.9, xtick=data,
	xticklabels from table={\mytable}{paradigm},
	xticklabel style={rotate=45,anchor=north east,inner sep=0mm},
	ylabel={\Large Time (s)}, ylabel near ticks]
	\addplot table [x expr=\coordindex+1,y=lio] {\mytable};
	\addlegendentry{\lioDeian};
	\addplot table [x expr=\coordindex+1,y=lioC] {\mytable};
	\addlegendentry{\lioHybrid};
	\addplot table [x expr=\coordindex+1,y=lioS] {\mytable};
	\addlegendentry{\lioStatic};
	\end{axis}
	\end{tikzpicture}
	\begin{tabular}{r|c|cc|cc|}
		& \lioDeian & \multicolumn{2}{c|}{\lioHybrid} & \multicolumn{2}{c|}{\lioStatic} \\ \cline{2-6} 
		& Time       & Time      & Speed-up     & Time      & Speed-up     \\ \hline
		\gls{datastar} (\S~\ref{bench:datastar})	&  10.571s         & 9.602s            &    9.1\%       &  9.229s          &  12.6\%   \\    
		\gls{busstar} (\S~\ref{bench:busstar})	&  7.144s         & 5.610s            &    21.4\%       &  3.238s           &  54.6\%   \\  
		\gls{mmustar} (\S~\ref{bench:mmustar})	&  16.179s         & 12.449s            &    23\%       &  8.450s           &  47.7\%   \\          
	\end{tabular}
	\caption{Performance measurement of all presented uses cases.}
	\label{table:benchmark_speed}
\end{figure}
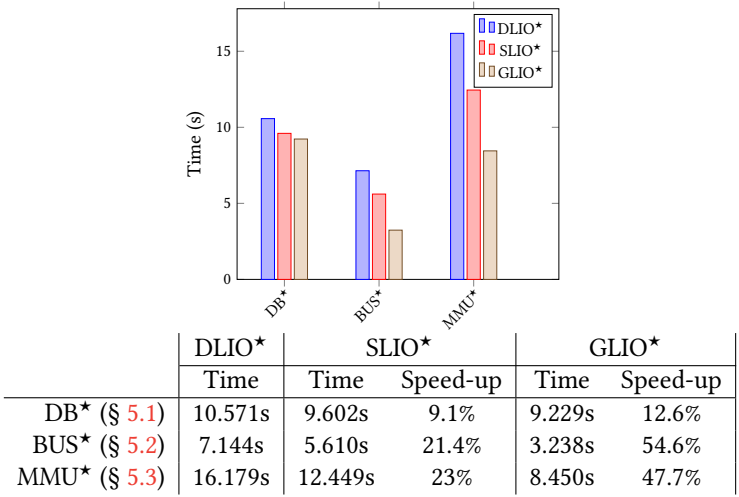

Figure~\ref{table:benchmark_size}
summarizes the \gls{lop} and the extracted Lines of C Code (LoC)
for each of our three benchmarks. 
Figure~\ref{table:benchmark_speed} summarizes the running times for each of the 
benchmark and the speedup of each library version with respect to 
\lioDeian.
To measure these times, we run
the \gls{busstar} and the \gls{mmustar} benchmarks $75M$ times 
and the \gls{datastar} benchmarks $20k$ times. 
Concerning data set, 
\gls{datastar} has a database with 2000 papers where 1992 belong to the user Mary,
the \gls{busstar} uses a list of packet that test all combination of the lattice labels, and
the \gls{mmustar} doesn't need a dataset.    

\begin{figure}[ht]
	\begin{small}
		\centering\begin{tabular}[t]{l r}
			\hline
			\gls{gcc} & 8.2.1 20181215  \\
			Fedora & 5.3.6-100.fc29.x86\_64 \\
			\gls{fstar} & \small{aeb4203c841735a6f401ed8b9cd44412a68c82bb} \\
			\gls{krml} & \small{882534387830a4cc259c1a543e0dfc10dcc70f52} \\ \\
			\hline
		\end{tabular}~\begin{tabular}[t]{l r}
			\hline
			Processor & Intel(R) Xeon(R) W-2104 \\
			Frequency & 3.2 GHz \\
			Core & 4 \\
			Architecture & x86\_64 \\
			RAM &  16GB \\
			\hline
		\end{tabular}
	\end{small}
	\caption{Details our benchmark platform}
	\label{table:hardware_used}\label{table:software_used}
\end{figure}

We conclude that moving 
from \lioDeian, 
to \lioHybrid, and
to \lioStatic, is expected to increase the  \gls{lop} but reduces the  \gls{loc}, 
since \gls{ifc} is statically checked. 
Since the lines of C code are reduced 
and runtime checks and label book keeping is removed, 
this leads to runtime speedups. 
From our three benchmarks we conclude that this 
extra effort in static verification pays off 
in low-level applications, like \gls{mmustar} and \gls{busstar}
that use few but critical \gls{ifc} checks. 
On the other hand, in applications like \gls{datastar} that are heavy 
on dynamic \gls{ifc} checks the verification effort is high and the 
effect on runtime improvement small. 

	\section{Related Works}

\gls{liostar} descends from a vast line of works started from the basis of MAC in \cite{bell1973secure} to the general lattice-theoretic model proposed by Denning in  \cite{denning1977certification} to verify the control information flow on computer or operating systems.

Today, large, security-sensitive, applications use expensive and state-of-the-art architectures, such as Risc-V or ARM designs, to implement systematic resource isolation at supervisor and hardware-MMU kernel level to safely sandbox legacy services of large corporate or cloud infrastructures. The open-source RISC-V architecture supports extensions providing hardware \gls{ifc} capabilities encoded as a byte-size tag alongside with data \cite{palmiero2018design,ferraiuolo2018hyperflow} to control data flow in accordance with the tags privileges.  ARM's Trustzone allows to segregate encrypted and decrypted data in physically isolated trust zones. \cite{de2015micro} generalizes this meta-data tag mechanism to implement more general software-defined \gls{ifc} policies at hardware level.

Virtualization technology and resource isolation available in modern operating systems and verified micro-kernels \citep{klein2009sel4,gu2016certikos}  is however far from available to consumer-market, IoT-oriented, embedded micro-controller architectures. On such targets, compartmentalization is a cost-effective compilation technique to complement label-enforced \gls{ifc} policy with defensive code to isolate possible software faults and prevent program threads from addressing data outside of their designated partitions \cite{de2015micro,10.1007/978-3-030-17184-1_18}.

Software-defined \gls{ifc} helps to overcome hardware limitations and can,  where available, strengthen coarse-grain, hardware security mechanisms (trust zones, virtualization, tags) with fine-grain user-, task- or channel-level micro-policies  \cite{de2015micro}. Software-level \gls{ifc} was first proposed in \cite{myers1999jflow} to annotate Java programs with \gls{ifc} policies.  \cite{hammer2006information} provides a language-agnostic library to check \gls{ifc} properties in imperative C or Java programs.

Dynamic \gls{ifc} policies have extensively been developed in operating system design.  \cite{zeldovich2006making} provides a  survey covering this domain.  For instance, \cite{krohn2007information} proposes operating system mechanisms to systematically check information flow read or written by system threads. \gls{lio} is introduced in \cite{giffin2012hails} and showed in \cite{Parker:2019:LIF:3302515.3290388} to support dynamic \gls{ifc} for web applications using a verified implementation in Liquid Haskell.  In \cite{buiras2015hlio}, \gls{lio} mixes static and dynamic verification in Haskell to mitigate the constraints of runtime checks. 

\gls{ifc}~\citep{Sabelfeld:2006} policies can be used to ensure component separation, but current techniques that enforce such policies either use heavy runtime checks~\citep{Austin12,Austin17,Yang16} or rely on advanced type-checking using high level programming languages~\citep{Schoepe14,buiras2015hlio}.

\gls{lio} relies on Haskell's monads to effectively enforce \gls{ifc} policies when reading/writing to databases or the web~\citep{stefan:2017:flexible,Parker:2019:LIF:3302515.3290388}.  \cite{gregersen2019dependently} presents the implementation of a statically verified \gls{ifc} policy in Idris. Idris is a pure functional programming language with dependent type and proof assistance. \cite{gregersen2019dependently} shows how modelling \gls{ifc} using dependent types improves expressibility of \gls{ifc} policies.

Hence, direct application of software-defined \gls{ifc} policies to embedded devices with, \eg \gls{lio}, faces two major obstacles. First, \gls{lio}'s policy enforcement relies on automatically generated runtime checks that would, if not properly sand-boxed, cause an unattended device to crash unpredictably.  Second, garbage collection and lazy evaluation in high-level languages may overflow their limited memory if implemented without extensive engineering, testing and profiling efforts. 

Our approach takes advantage of both the expressivity of dependent-types in the verified programming language \gls{fstar}~\citep{mumon}, allowing us to use \gls{fstar}'s effects to encode monadic \gls{ifc} encapsulation, and the capability of generating bare-metal system code, by using its \gls{krml}~\citep{lowstar} code generator. This approach yields three advantages: 

Like related approaches based on high-level programming languages, \citep{stefan:2017:flexible,Parker:2019:LIF:3302515.3290388,gregersen2019dependently,buiras2015hlio}, \gls{liostar} lifts policy enforcement from runtime checks to static proof obligations by using powerful dependent-type systems using a minimalistic effect system defined from \gls{fstar}'s Dijkstra Monads~\citep{Maillard19}, leading to verified code with minimal mechanisation, and in similar ways to \citep{Parker:2019:LIF:3302515.3290388}, using Liquid Haskell's refinement type system.

\cite{buiras2015hlio} offers a different hybridation mechanism that ours: it eliminates \gls{ifc} runtime checks that can be ruled safe statically and keep other, call-dependent, dynamic checks.  This is a most suitable approach for transactional applications, where throwing an exception from some \gls{lio} client application is non-critical or fail-safe.  However, in the case of, possibly unattended, reactive applications, this is  not an option, as failing safe usually means to restart a real-time and potentially mission- or safety-critical application.

\cite{10.1007/978-3-642-11486-1_30} provides a detailed review on the extensive number of related approaches based on the static analysis of imperative system programs. The recent \cite{Guarnieri20}, for instance, statically analyses bytecode to monitor programs that may leak unintended information when executed on speculative architectures.  As in these approaches,  \gls{liostar} offers the capability to run verified code generated from the \gls{krml} compiler ~\citep{lowstar}, without the need for a runtime library or a garbage collector, and hence for direct application for low-level, resource-constrained, embedded architectures.  

\cite{gregersen2019dependently} models \gls{lio} in the domain-specific language Idris to mechanically verify its specification by theorem proving; and \cite{Parker:2019:LIF:3302515.3290388} uses the refinement reflection mechanism of Liquid Haskell \cite{10.1145/3158141} to automate the proof  for a model of \gls{lio} in the $\lambda$-calculus.   \gls{liostar} verifies the actual \gls{liostar}'s client applications correct and non-interfering by using \gls{fstar}'s meta-programming environment \gls{metastar}~\citep{metafstar}.   

\cite{10.1007/978-3-642-11486-1_30} shows static verification of \gls{ifc} policies to be as strong as its, more permissive, dynamic enforcement via defensive code or monitors, both static and dynamic approaches ultimately providing the same assurance of termination-insensitive non-interference.

	\section{Conclusion and Future Works}
We presented \gls{liostar} a verified \gls{fstar}, \gls{ifc} framework.
First, we implemented three library versions: 
1) the dynamic \lioDeian, where \gls{ifc} policies are checked at runtime, 
2) the static \lioHybrid, where \gls{ifc} policies are lifted to compile time proof obligations, and 
3) the ghost \lioStatic, where the \gls{ifc} label tracking 
is totally erased at runtime. 
Next, we used metaprogramming 
to define |genNILemma|, a procedure that 
generates a lemma stating  
that a concrete \lioStatic client is noninterferent. 
We applied |genNILemma| to 
two of our benchmarks and various smaller examples, 
and generated noninterference lemmata that 
\gls{fstar} automatically proved correct. 
We evaluated our approach using three benchmarks
and validated that moving 
from \lioDeian, to \lioHybrid, to \lioStatic,
the verification effort (\ie the \gls{fstar} Lines of Code) 
increases, but the extracted C code is cleaner (and shorter)
and up to \%54.6 faster (in the case of \gls{busstar}). 
In general, we propose a methodology that is using 
static verification to reduce runtime checks 
and metaprogramming to prove program metaproperties,
leading to both fast and provably correct software, 
ideal to be executed by embedded devices.

\subsection*{Future works}

\paragraph{Metaproperties} 

To verify our methodology, we developed a mechanized noninterference proof 
of \lioStatic's clients. 
As discussed in~\S~\ref{subsec:metatheory:limitations} our approach
suffers from various limitations, some of them stem from \gls{metastar}, 
which prevented us from proving noninterference of the \acrshort{datastar} clients. 
As~\gls{metastar} gets more mature and our experience with it grows, 
we aim to complete the metatheory of \acrshort{datastar} in the near future 
and even apply this metaprogramming for metaproperties methodology to 
generalize a noninterference proof for every \lioStatic client. 

\paragraph{Layered Effects}

Our current implementation of \gls{liostar} suffers from a lack of parametricity: 
it is not possible to use our framework without, for instance, \gls{lowstar}'s |ST| effect,
the standard \gls{fstar} way of writing low-level programs.
Layered effect, a new \gls{fstar} feature, addresses this issue.

This last improvement of \gls{fstar} will enable us to 
define \gls{ifc} effects independently of target IO 
computation (memory, streams, etc) and allow to express \gls{ifc}
policies in a much simpler and elegantly modular way by
the composition of effects.
This will make our \gls{ifc} library a lot more
portable and allow to significantly reduce its \gls{tcb}. 

Our current implementation of \gls{liostar} only handles pure functions 
and we couldn't yet use \gls{lowstar} effects to extend it with a complete 
memory model, which would have been much of an improvement for, 
\eg the \gls{mmustar} case study. 
There are several reasons why we made the design choice to limit our 
working prototype with pure effects: 
1/ using \gls{fstar}'s base effects would have implied modifying the 
\gls{lowstar} API to include labels, 
2/ furthermore, the \gls{lowstar} memory model and API are going to be revamped 
(C code extraction from \gls{krml} doesn't currently play nicely with reifiable effect), 
and 3/ the notion of layered effect, a new way of compositionally declaring effects, 
is currently being tested in \gls{fstar}.

This last improvement of \gls{fstar} will enable us to define \gls{ifc} 
effects independently of target IO monad (memory, streams, etc) 
and allow to express \gls{ifc} policies in a much simpler and elegantly modular 
way by the composition of effects. 
This will make our \gls{ifc} library a lot more portable and allow us to significantly reduce its \gls{tcb}.

\bibliographystyle{ACM-Reference-Format}
\bibliography{reference} 

%
%
%
%

\end{document}